\documentclass[preprint,aps,showpacs,nofootinbib]{revtex4}

\usepackage{epsfig,amssymb,amsmath,cancel}

\def\bea{\begin{eqnarray}}
\def\eea{\end{eqnarray}}
\def\beq{\begin{equation}}
\def\eeq{\end{equation}}

\begin{document}

\draft \tighten

\preprint{KIAS-P08085}
\preprint{TU-835}
\title{\large \bf Sparticle masses in deflected mirage mediation}
\author{
Kiwoon Choi\footnote{email: kchoi@muon.kaist.ac.kr}$^1$,
Kwang Sik Jeong\footnote{email: ksjeong@kias.re.kr}$^2$ \\
Shuntaro Nakamura\footnote{email: shuntaro@tuhep.phys.tohoku.ac.jp}$^3$,
Ken-Ichi Okumura\footnote{email: okumura@tuhep.phys.tohoku.ac.jp}$^3$,
Masahiro Yamaguchi\footnote{email: yama@tuhep.phys.tohoku.ac.jp}$^3$}
\affiliation{
$^1$Department of Physics, KAIST, Daejeon 305-701, Korea \\
$^2$School of Physics, Korea Institute for Advanced Study,
Seoul 130-722, Korea \\
$^3$Department of Physics, Tohoku University, Sendai 980-8578, Japan
}


\vspace{2cm}

\begin{abstract}
We discuss the sparticle mass patterns that can be realized in deflected
mirage mediation scenario of supersymmetry breaking, in which the moduli,
anomaly, and gauge mediations all contribute to the MSSM soft parameters.
Analytic expression of low energy soft parameters and also the sfermion
mass sum rules are derived, which can be used to interpret the experimentally
measured sparticle masses within the framework of the most general mixed
moduli-gauge-anomaly mediation.
Phenomenological aspects of some specific examples are also discussed.
\end{abstract}

\maketitle

\section{Introduction}

Weak scale supersymmetry (SUSY)  is one of the prime candidates for physics
beyond the standard model at the TeV scale \cite{Nilles:1983ge}.
Low energy phenomenology of weak scale SUSY is determined mostly by the soft
SUSY breaking terms of the visible gauge and matter superfields.
Those soft terms are required to preserve flavor and CP with a good accuracy,
which severely constrains the possible mediation mechanism of SUSY breaking.
Presently, there are three known mediation schemes to yield flavor and CP
conserving soft terms\footnote{Even in these schemes, there can be dangerous
CP violation from the Higgs $\mu$ and $B$ parameters, which should be
considered separately.}: gauge mediation \cite{gauge.mediation.1,gauge.mediation.2},
anomaly mediation \cite{anomaly.mediation}, and string dilaton or volume-moduli
mediation \cite{dilaton.mediation}.
In gauge and anomaly mediations, the radiative corrections due to the standard
model (SM) gauge interaction play dominant role for the mediation, and thereby
the resulting soft terms automatically preserve flavor and CP.
For dilaton/moduli mediation, soft terms induced by the dilaton/moduli
$F$-components preserve flavor and CP by different reasons.
The couplings between the messenger dilaton/moduli and the MSSM matter fields
preserve flavor as they are determined by family-universal rational numbers
called the modular weight \cite{dilaton.mediation}, and preserve CP as
a consequence of the associated axionic shift symmetries \cite{Choi:1993yd}.

So far, most studies of SUSY phenomenology have been focused on the cases
that SUSY breaking is dominated by one of the dilaton/moduli,
gauge and anomaly mediations. However, recent progress in moduli stabilization
suggests that it is a rather plausible possibility that moduli mediation and
anomaly mediation are comparable to each other
\cite{Choi:2004sx,Endo:2005uy,Choi:2005uz}, which can be naturally realized
in KKLT-type moduli stabilization scenario \cite{Kachru:2003aw}.
The resulting soft terms show a distinct feature that sparticle masses are
unified at a mirage messenger scale hierarchically lower than the scale of
gauge coupling unification \cite{Choi:2005uz}.
Also a mixed scheme of  anomaly and gauge mediations has been proposed before
as a solution to the tachyonic slepton problem of anomaly mediation
\cite{Pomarol:1999ie}.
Recently, it has been pointed out that these schemes can be generalized to
include the three known flavor and CP conserving mediations altogether
\cite{Nakamura:2008ey,Everett:2008qy}.
Such a most general mixed mediation has been dubbed `deflected mirage mediation'
as sfermion masses are deflected from the mirage unification trajectory due to
the presence of gauge mediation.

In this paper, we wish to examine in more detail the sparticle mass pattern
in deflected mirage mediation scenario, together with some phenomenological
aspects of the scheme.
The organization of this paper is as follows.
In section II, we discuss a class of string-motivated effective supergravity
models that realize deflected mirage mediation scenario.
In section III, we analyze the renormalization group running of soft parameters
to derive the (approximate) analytic expression of low energy sparticle masses
in deflected mirage mediation, which can be used to interpret the experimentally
measured sparticle masses within the framework of the most general flavor and CP
conserving mediation scheme.
We discuss in section IV the phenomenological feature of two specific examples,
one with an accidental little hierarchy between $m^2_{H_u}$ and other soft
mass-squares and another with gluino NLSP, that can be obtained within
deflected mirage mediation scenario.
Section V is the conclusion.

\section{Effective supergravity for deflected mirage mediation}

In this section, we discuss a class of  4-dimensional (4D) N=1 supergravity
(SUGRA) models that realize the deflected mirage mediation scenario.
The models discussed here may arise as a low energy effective theory of
KKLT-type flux compactification or its variants in string theory.
The model contains first of all the MSSM gauge and matter superfields, $V^a$
and $Q_i$, and also vector-like MSSM-charged exotic matter superfields,
$\Phi+\Phi^c$, which live on the visible sector brane.
There are light moduli $T_I$, e.g. the K\"ahler moduli, stabilized by
non-perturbative effects encoded in the superpotential, and also heavy moduli
$U_p$ stabilized by flux, e.g. the complex structure moduli.
Typically ${\rm Im}(T_I)$ corresponds to an axion-like field, and thus the
couplings of $T_I$ are invariant under the axionic shift symmetry:
\bea
\label{axionicshift}
U(1)_{T_I}:\, T_I\rightarrow T_I
+ \mbox{imaginary constant},
\eea
upon ignoring  exponentially small non-perturbative effects.

In KKLT-type compactification, moduli stabilization dynamics itself does not
break SUSY since the flux and non-perturbative effects stabilize moduli at
a supersymmetric AdS vacuum.
Thus, to break SUSY and lift the vacuum to dS state, one needs to introduce
a SUSY breaking brane separately.
This SUSY breaking brane might be an anti-brane that exists in the underlying
string theory, or a brane carrying a 4D dynamics that breaks SUSY
spontaneously.
An important feature of KKLT-type  compactification is that it involves
a highly warped throat produced by flux.
In the presence of such a warped throat, SUSY breaking brane is stabilized
at the tip of throat where the potential of the position modulus is minimized.
On the other hand, to implement the high scale gauge coupling unification
in the MSSM, the visible sector brane should be stabilized within the
internal space at the UV end of throat.
This results in a warped separation between the visible brane and the SUSY
breaking brane, making the visible sector and the SUSY breaking sector to be
sequestered from each other \cite{warped_sequestering}.
To be specific, here we will consider a SUSY braking sector described by
a Polony-like superfield $Z$ having a linear superpotential.
However, it should be stressed that the visible sector soft terms which are
of our major concern are independent of how SUSY is broken at the tip of
throat, and therefore  our subsequent discussion is valid in cases that SUSY
is broken by other means, e.g. by an anti-brane \cite{Choi:2006bh}.

\subsection{Effective supergravity action}

With the above features of KKLT-type compactification, the 4D effective
action can be written as
\bea
\label{4daction}
{\cal L}_{4D}&=& \int d^4\theta\, CC^*\Big[\Omega_{\rm mod}
+ \Omega_{\rm matter} + \Omega_{\rm polony}\,\Big]
\nonumber\\
&&
+\, \left[\,\int
d^2\theta \, \Big(\, \frac{1}{4}f_aW^{a\alpha}W^a_\alpha
+ C^3\Big\{W_{\rm mod} + W_{\rm matter}+W_{\rm polony}\Big\}
+ {\rm h.c.}\,\right],
\eea
where
\bea
\Omega_{\rm mod} &=& \Omega_{\rm mod}(U_p,U_p^*,T_I+T_I^*),
\nonumber \\
W_{\rm mod}&=&W_{\rm flux}(U_p)+\sum_I
A_I(U_p)e^{-8\pi^2a_IT_I},
\nonumber
\eea
for a flux-induced superpotential $W_{\rm flux}(U_p)$ and  the
nonperturbative term $e^{-8\pi^2 a_IT_I}$  with real parameter $a_I$ of
order unity,
\bea
\label{matter-functions}
\Omega_{\rm matter}&=& \sum_A {\cal Y}_A(U_p,U_p^*,T_I+T_I^*)\Phi^{A*}\Phi^A
\quad \left(\Phi^A=Q_i, \Phi, \Phi^c, X\right),
\nonumber \\
W_{\rm matter} &=&
\lambda_\Phi(U_p)X\Phi\Phi^c+\frac{\kappa(U_p)X^n}{M_{Pl}^{n-3}}
+ \frac{1}{6}\lambda_{ijk}(U_p)Q_iQ_jQ_k,
\eea
where $Q_i$ are the MSSM matter superfields, $\Phi+\Phi^c$ are exotic
vector-like matter superfields, and $X$ is a singlet superfield giving a mass
to $\Phi+\Phi^c$, and
\bea
\Omega_{\rm polony}&=&ZZ^*-\frac{(ZZ^*)^2}{4M_*^2},
\nonumber \\
W_{\rm polony}&=&M_{\rm SUSY}^2Z,
\eea
for a Polony-like field $Z$ which breaks SUSY at the tip of throat.
Here $C$ is the chiral compensator superfield, $f_a$ are the gauge kinetic
functions of the MSSM gauge fields, and we are using the SUGRA unit with
$M_{Pl}=1$, where $M_{Pl}=\sqrt{G_N/8\pi}\simeq 2\times 10^{18}\,{\rm GeV}$.

As $Z$ is localized at the tip of throat, and thus is sequestered from the
visible sector \cite{warped_sequestering}, there are no contact interactions
between $Z$ and the visible sector fields in the superspace action, which
means ${\cal Y}_A$, $f_a$, $\lambda_{ijk}$, $\lambda_\Phi$, and $\kappa$ are
all independent of $Z$.
Also the axionic shift symmetry (\ref{axionicshift}) requires that ${\cal
Y}_A$ is a function of the invariant combination $T_I+T_I^*$, the holomorphic
couplings $\lambda_{ijk}, \lambda_\Phi$ and $\kappa$ are independent of $T_I$,
and $\partial f_a/\partial T_I$ are real constants.
To incorporate the anomaly mediated SUSY breaking, one needs to include
the logarithmic $C$-dependence of $f_a$ and ${\cal Y}_A$, which is associated
with the renormalization group (RG) running of the gauge and Yukawa couplings.
Then, under the constraints from the axionic shift symmetry, $f_a$ and
${\cal Y}_A$ can be written as
\bea
f_a&=&\tilde{f}_a(T_I,U_p)+\frac{b_a}{16\pi^2}\ln C= \sum_I
k_IT_I+\epsilon(U_p)+\frac{b_a}{16\pi^2}\ln C,\nonumber
\\
\ln {\cal Y}_A&=& \ln\tilde{\cal
Y}_A(T_I+T_I^*,U_p,U_p^*)+\frac{1}{8\pi^2}\int^{\mu/\sqrt{CC^*}}_{M_{GUT}}
\frac{d\mu^\prime}{\mu^\prime}\gamma_A,
\eea
where $b_a$ and $\gamma_A$ are the one-loop beta function coefficient and the
anomalous dimension of $\Phi^A$, respectively, and $k_I$ are real parameters
of order unity.
Here we assume the gauge coupling unification around the scale
$M_{GUT}\simeq 2\times 10^{16}\,{\rm GeV}$, which requires $k_I$ to be universal for
the MSSM gauge kinetic functions $f_a$. In fact, as $f_a$ corresponds to an
Wilsonian gauge coupling, the one-loop coefficient of $\ln C$ in $f_a$
depends on the corresponding regularization scheme.
On the other, the 1PI gauge coupling does not have such scheme dependence.
Here we have chosen a specific scheme that the one loop $C$-dependence of the
1PI gauge coupling is fully encoded in the Wilsonian coupling, for which
$b_a$ is given by the one-loop beta function coefficient.

As for the stabilization of $X$, one can consider two scenarios.
The first scenario is that $X$ is stabilized by the combined effects of
the SUSY breaking by $F^C$ and the non-renormalizable operator
$\kappa X^n/M_{Pl}^{n-3}$ ($n >3$).
Another possibility is that $\kappa=0$, and $X$ is stabilized by the radiative
correction to its K\"ahler potential.
In fact, both scenarios give a similar size of $F^X/X$, while the resulting
mass of $X$ is quite different.
In the first scenario, all components of $X$ get a mass comparable to the
gravitino mass which is of ${\cal O}(10)\,{\rm TeV}$ \cite{Pomarol:1999ie}.
On the other hand, in the second scenario dubbed `axionic mirage mediation',
the pseudo-scalar component of $X$ can be identified as the nearly
massless QCD axion solving the strong CP problem, and its fermionic partner,
the axino,  gets a  two-loop suppressed small mass relative to the gravitino
mass \cite{Nakamura:2008ey}.

As for the SUSY breaking sector, we have taken a simple example given by
\bea
\Omega_{\rm polony}&=&ZZ^*-\frac{(ZZ^*)^2}{4M_*^2},
\nonumber \\
W_{\rm polony}&=&M_{\rm SUSY}^2Z,
\eea
where $M_{\rm SUSY}$ and $M_*$ are the two mass parameters for SUSY-breaking
dynamics.
Generically, some moduli may have a non-negligible wavefunction value at
the tip of throat.
Then those moduli can have a sizable contact interaction with $Z$, which means
that $M_{\rm SUSY}$ and $M_*$, as well as the coefficient of $ZZ^*$ in
$\Omega_{\rm polony}$, become a nontrivial function of moduli.
However, such an additional complexity does not affect our subsequent
discussion, and thus here we consider the simple case that $M_{\rm SUSY}$
and $M_*$ are moduli-independent constants.
At any rate, with the above form of $\Omega_{\rm polony}$ and $W_{\rm polony}$,
the vacuum value of $Z=z+\theta\tilde{z}+F^Z\theta^2$ is determined as
\bea
\langle Z\rangle =-\frac{C_0^{*2}}{C_0}M_{\rm SUSY}^{2}\theta^2,
\eea
where $C_0$ is the scalar component of the compensator superfield $C$.
The scalar component $z$ gets a mass
\bea
m_{z}\,\sim \, M_{\rm SUSY}^2/M_*,
\eea
while the fermion component $\tilde{z}$ corresponds to the Goldstino.
Due to the warping, both $M_{\rm SUSY}$ and $M_*$ are red-shifted by an
exponentially small warp factor at the tip of throat:
\bea
M_{\rm SUSY}\,\sim \,M_* \,\sim\, e^{-A}
M_{Pl},\eea where \bea \left.g_{\mu\nu}\right|_{\rm tip}=e^{-2A}
\eta_{\mu\nu}.
\eea

\subsection{Integrating out heavy moduli and Polony-like field}

The flux-induced superpotential of $U_p$ can be expanded around its stationary
point:
\bea
W_{\rm flux}(U_p)&=& W_{\rm flux}(\tilde{U}_p)+\frac{1}{2}
\frac{\partial^2 W_{\rm flux}(\tilde{U})}{\partial U_p\partial
U_q}(U_p-\tilde{U}_p)(U_q-\tilde{U}_q)+\cdots
\nonumber \\
&\equiv& w_0+\frac{1}{2}\left(M_U\right)_{pq}
(U_p-\tilde{U}_p)(U_q-\tilde{U}_q)+ \cdots,
\eea
where $\tilde{U}_p$ denotes the stationary point of $W_{\rm flux}$:
\bea
\left.\frac{\partial W_{\rm flux}}{\partial U_p}\right|_{U_q=\tilde{U}_q}=0.
\eea
Due to the quantization of flux, for generic flux configuration, both $w_0$
and $M_U$ would be of order unity in the unit with $M_{Pl}=1$.
However, if SUSY breaking is initiated at the tip of throat with a red-shifted
$M_{\rm SUSY}\sim e^{-A}M_{Pl}$, we are required to consider a special type
of flux configuration giving an exponentially small
\bea
w_0=\left.{W}_{\rm flux}\right|_{U_p=\tilde{U}_p}\sim e^{-2A},
\eea
in order to get a nearly vanishing vacuum energy density schematically given by
\bea
V_{\rm vac}=|M_{\rm SUSY}|^4-|w_0|^2.
\eea
Still the flux-induced moduli mass matrix $M_U$ generically has the eigenvalues
of order unity.
Therefore, in flux compactification scenario with SUSY breaking initiated at the
tip of throat, one has the mass hierarchy:
\bea
\label{warpedhierarchy}
m_{3/2} \sim e^{-A} m_z \sim e^{-2A}M_U,
\eea
where $m_{3/2}$ is the gravitino mass, $M_U$ denotes the supersymmetric mass of
the flux-stabilized moduli $U_p$, and $m_z\sim M_{\rm SUSY}^2/M_*$ is the
non-supersymmetric mass of the scalar component of the Polony-like superfield
localized at the tip of throat.

With the mass hierarchy (\ref{warpedhierarchy}), we can integrate out $U_p$ and
$z$ to construct the effective theory of light fields including the visible
sector fields $\Phi^A=(Q_i,\Phi,\Phi^c,X)$, chiral compensator $C$, light moduli
$T_I$, and the Goldstino $\tilde{z}$.
This can be done by solving the following superfield equations of motion within
the expansion in powers of the warp factor $e^{-A}$:
\bea
&& \frac{1}{4}\bar{\cal D}^2\left(CC^*
\frac{\partial \Omega_{\rm mod}}{\partial U_p}\right)
+ C^3\frac{\partial W_{\rm mod}}{\partial U_p}=0,
\nonumber \\
&& \frac{1}{4}\bar{\cal D}^2\left(CC^*
\frac{\partial \Omega_{\rm polony}}{\partial Z}\right)
+ C^3\frac{\partial W_{\rm polony}}{\partial Z}=0,
\eea
where $\bar{\cal D}^2=\bar{\cal D}^{\dot{\alpha}}\bar{\cal D}_{\dot{\alpha}}$
is the superspace covariant derivative. It is straightforward to find that the
solutions are given by
\bea
\label{solution}
U^{\rm sol}_p &=& \tilde{U}_p+{\cal O}\left(\frac{\bar{\cal D}^2}{M_U},
\frac{m_{3/2}}{M_U}\right),
\nonumber \\
Z^{\rm sol} &=& -\frac{C^{*2}}{C}M_{\rm SUSY}^2\Lambda^2
+ {\cal O}\left(\frac{\bar{\cal D}^2}{m_z},\frac{m_{3/2}}{m_z}\right),
\eea
where $\Lambda^\alpha$ is the Goldstino superfield defined as
\bea
\Lambda^\alpha=\theta^\alpha
+ \frac{1}{M_{\rm SUSY}^2}\tilde{z}^\alpha+\cdots,
\eea
with the ellipsis denoting the Goldstino-dependent higher order terms.
Note that $\bar{\cal D}^2$ acting on light field eventually gives rise to an
$F$-component which is of  ${\cal O}(m_{3/2})$ up to a factor of
${\cal O}(8\pi^2)$.

One can now derive the effective action of light fields by replacing $U_p$
and $Z$ with $U_p^{\rm sol}$ and $Z^{\rm sol}$.
At leading order in $e^{-A}$, the effective action is obtained by simply
replacing $U_p$ with $\tilde{U}_p$, and $Z$ with
$-M_{\rm SUSY}^{2}\Lambda^2C^{*2}/C$:
\bea
{\cal L}_{\rm eff}=\left.{\cal L}_{4D}\right|_{U_p=\tilde{U}_p,
Z=-M_{\rm SUSY}^2\Lambda^2C^{*2}/C}.
\eea
The resulting effective action is given by
\bea
\label{effective}
{\cal L}_{\rm eff}&=& \int d^4\theta \,
\Big[\,-3CC^*e^{-K_{\rm eff}/3}
- C^2C^{*2}M_{\rm SUSY}^4\Lambda^2\bar{\Lambda}^2\,\Big]
\nonumber \\
&& +\,\Big[\int d^2\theta\, \Big(\,\frac{1}{4}f^{\rm eff}_a
W^{a\alpha}W^a_\alpha +C^3W_{\rm eff}\,\Big)+{\rm h.c.}\,\Big],
\eea
where
\bea
\label{effective1}
K_{\rm eff}&=&K_0(T_I+T_I^*)+{\cal Z}_A(T_I+T_I^*)\Phi^A\Phi^{A*},
\nonumber \\
f^{\rm eff}_a&=& \tilde{f}_a(T_I)+\frac{b_a}{16\pi^2}\ln C=
\sum_I k_IT_I +\epsilon +\frac{b_a}{16\pi^2}\ln C,
\nonumber \\
W_{\rm eff}&=&w_0+\sum_I A_Ie^{-8\pi^2 a_I T_I}
+ \frac{\kappa X^n}{M_{Pl}^{n-3}}+\lambda_\Phi
X\Phi\Phi^c+\frac{1}{6}\lambda_{ijk}Q_iQ_jQ_k,
\eea
for the moduli K\"ahler potential $K_0$ and the matter K\"ahler metric
${\cal Z}_A$ determined as
\bea
-3e^{-K_0/3}&=&\left.\Omega_{\rm mod}(U_p,U_p^*,T_I+T_I^*)
\right|_{U_p=\tilde{U}_p},
\nonumber \\
e^{-K_0/3}{\cal Z}_A&=&\left.{\cal Y}_A(U_p,U^*_p,\ln(CC^*),T_I+T_I^*)
\right|_{U_p=\tilde{U}_p}.
\eea
The above effective action is defined at a renormalization point $\mu$ below
the compactification scale, but above the mass of the exotic matter field
$\Phi+\Phi^c$.
Note that this renormalization point can be higher than the masses of the
integrated heavy moduli $U_p$ and Polony scalar $z$, while it should be lower
than the mass scale characterizing the non-renormalizable interactions between
the integrated fields and the remained light fields, which is of order the
compactification scale or the GUT scale in our case.
In the procedure to integrate out the Polony scalar to obtain the Akulov-Volkov
action of the Goldstino superfield, we have used the following identity for the
Goldstino superfield in the flat spacetime limit \cite{Samuel:1982uh}:
\bea
\frac{1}{4}\bar{\cal D}^2(\Lambda^2\bar{\Lambda}^2)=
-\Lambda^2\left(1-2i\partial_\mu
\Lambda\sigma^\mu\bar{\Lambda}-4\bar{\Lambda}^2\partial_\mu\Lambda
\sigma^{\mu\nu}\partial_\nu\Lambda\right),
\eea
and ignored the higher order Goldstino operators as well as the higher
derivative operators.

\subsection{Supersymmetry breaking}

In the class of models discussed here, the SUSY breaking field $Z$ is
sequestered from the visible sector, and then the MSSM soft terms are
determined by $F^C$, $F^{T_I}$ and $F^X$, which characterize the anomaly,
moduli, and gauge mediation, respectively.
As we have noticed, in order to get a nearly vanishing cosmological constant
with $M_{\rm SUSY}/M_{Pl}\sim e^{-A}$, one needs a special type of flux
configuration yielding $m_{3/2}/M_{Pl}\sim w_0/M_{Pl}^3\sim e^{-2A}$.
On the other hand, nonperturbative dynamics generating the superpotential
term $A_Ie^{-8\pi^2 a_IT_I}$ originates from the UV end of throat, and
thus there is no significant red-shift for $A_I$.
This suggests that $A_I$ are generically of order unity in the unit with
$M_{Pl}=1$, and then
\bea
\ln (A_I/w_0)\,\simeq\, \ln(M_{Pl}/m_{3/2})\,\sim\, 4\pi^2
\eea
for $m_{3/2}={\cal O}(10)\,{\rm TeV}$.
In the presence of such a big hierarchy between $w_0$ and $A_I$, much of
the physical properties of $T_I$ and $X$ can be determined without knowing
the explicit form of their K\"ahler potential.
For instance, $T_I$ are stabilized near the supersymmetric solution of
$\partial_IW+(\partial_IK)W=0$ with a mass
\bea
m_{T_I}\sim m_{3/2}\ln(M_{Pl}/m_{3/2}).
\eea
If $\kappa\neq 0$ for some $n>3$, $X$ is stabilized at an intermediate scale
with
\bea
m_X\sim m_{3/2}.
\eea
The resulting vacuum expectation values of the scalar and $F$ components of
$T_I$ and $X$ (in the Einstein frame) are given by
(see Refs.\cite{Choi:2006za,Everett:2008qy} for explicit derivations)
\bea
\label{vev}
a_I T_I &\simeq&  \frac{\ln(M_{Pl}/m_{3/2})}{8\pi^2},
\nonumber \\
X &\sim& \left(\frac{M_{Pl}^{n-3}m_{3/2}}{\kappa}\right)^{1/(n-2)},
\nonumber \\
\frac{F^{T_I}}{T_I+T_I^*}&\simeq&
\frac{1}{\ln(M_{Pl}/m_{3/2})}\frac{F^C}{C},
\nonumber \\
\frac{F^X}{X}&\simeq& -\frac{2}{n-1}\frac{F^C}{C},
\eea
where
\bea
\frac{F^C}{C} &=& m^*_{3/2}+\frac{1}{3}F^P\partial_PK,
\nonumber \\
F^P &=& -e^{K/2}K^{P\bar{Q}}\left(\partial_QW+(\partial_QK) W\right)^*
\eea
for $\Phi^P=(T_I, X)$, and we have used
$\ln(A_I/w_0)\simeq \ln(M_{Pl}/m_{3/2})$.

As is well known, $F^C$ generates the anomaly-mediated soft parameters of
${\cal O}\left({F^C}/8\pi^2\right)$ \cite{anomaly.mediation} at a high
messenger scale around the compactification scale, while $F^{T_I}$ generates
moduli-mediated soft parameters of ${\cal O}\left(F^{T_I}\right)$
\cite{dilaton.mediation} at a similar high messenger scale.
In addition to these, the exotic vector-like matter fields $\Phi+\Phi^c$ give
rise to a gauge-mediated contribution of
${\cal O}\left(F^{M_\Phi}/8\pi^2 M_\Phi\right)$
\cite{gauge.mediation.1,gauge.mediation.2} at
the messenger scale $M_\Phi$, where  $M_\Phi$ and $F^{M_\Phi}$ denote the
scalar component and the $F$ component, respectively, of the messenger mass
given by
\bea
\label{mass_superfield}
\int d^2\theta \,C^3\Big(M_\Phi+\theta^2F^{M_\Phi}\Big) \Phi\Phi^c.
\eea
In the SUGRA models of the form (\ref{effective1}), $\Phi+\Phi^c$ get a mass
through the superpotential coupling $\lambda_\Phi X\Phi\Phi^c$, and then
\bea
\frac{F^{M_\Phi}}{M_\Phi}=\frac{F^X}{X}=
-\frac{2}{n-1}\frac{F^C}{C}\quad\mbox{with}
\,\,\,M_\Phi=\lambda_\Phi\langle X\rangle.
\eea
The most interesting feature of these SUGRA models is that
\bea
\frac{F^{T_I}}{T_I+T_I^*}\,\sim\,
\frac{1}{8\pi^2}\frac{F^C}{C}\,\sim\,
\frac{1}{8\pi^2}\frac{F^X}{X},
\eea
independently of the K\"ahler potential.
As a result, the MSSM soft parameters receive a similar size of contribution
from all of the moduli, anomaly, and gauge mediations.

Another interesting feature  is that  the phases of $F$ components are
dynamically aligned to each other as
\bea
{\rm arg}\left[\frac{F^C}{C}\left(\frac{{F}^{T_I}}{T_I+T_I^*}\right)^*
\right]=
\mbox{arg}\left[\frac{F^C}{C}\left(\frac{{F}^{X}}{X}\right)^*\right]=0.
\eea
With this feature, soft terms preserve CP, although they receive a comparable
contribution from three different origins.
For this dynamical alignment, the axionic shift symmetry (\ref{axionicshift})
plays an essential role \cite{Choi:1993yd,Choi:2004sx,Endo:2005uy}.
To see this, let us note that one can always make $w_0$ in the superpotential
to be real by an appropriate $U(1)_R$ transformation of the Grassmann
variables, make $A_I$ real by an axionic shift of $T_I$, and finally make
$\kappa$ real by a phase rotation of $X$, under which the K\"ahler potential
is invariant.
In this field basis, it is straightforward to see that ${\rm Im}(T_I)$ and
${\rm arg}(X)$ are stabilized at a CP conserving value, and therefore $W$,
$e^{-8\pi^2 a_IT_I}$, and $ X$ have real vacuum values.
As the K\"ahler potential is invariant under the axionic shift symmetry
(\ref{axionicshift}) and the phase rotation of $X$, the resulting vacuum
values of $\frac{F^C}{C}$, $\frac{F^X}{X}$, and ${F^{T_I}}$ are all real.

In deflected mirage mediation, soft parameters can preserve flavor in a natural
way.
To satisfy the FCNC constraints, the following moduli-mediated sfermion masses
and $A$-parameters are required to be (approximately) family-independent:
\bea
\tilde{m}_i^2&=&-F^{T_I}F^{T_J*}\partial_{T_I}\partial_{T^*_J}\ln
\left(e^{-K_0/3}{\cal Z}_i\right),
\nonumber \\
\tilde{A}_{ijk}&=&F^{T_I}\partial_{T_I}
\ln \left(e^{-K_0}{\cal Z}_i{\cal Z}_j {\cal Z}_k\right),
\eea
where ${\cal Z}_i$ is the K\"ahler metric of the MSSM matter field $Q_i$.
At leading order in the string coupling $g_{st}$ or the string slope parameter
$\alpha^\prime$, the $T_I$-dependence of ${\cal Z}_i$ is typically given by
\cite{dilaton.mediation}
\bea
\label{modular-weight}
{\cal Z}_i =\prod_I (T_I+T_I^*)^{n_I^i},
\eea
where $n_I^i$ is the modular weight of $Q_i$.
If different families with the same gauge charges originate from the same type
of branes or brane intersections, which is indeed the case in most of
semi-realistic string models, the matter modular weights are
{\it family-independent} rational numbers
\cite{Choi:2004sx,Conlon:2006tj,Choi:2006za,Choi:2008hn}, for which the
resulting $\tilde{m}_i^2$ and $\tilde{A}_{ijk}$ are family-independent.

In the above, we have considered the models of deflected mirage mediation,
in which the gauge messengers get a mass through the superpotential coupling
$\lambda_\Phi X\Phi\Phi^c$ with $X$ stabilized at an intermediate scale, either
by radiative effects or by the higher dimensional operator
$\kappa X^n/M_{Pl}^{n-3}$ ($n>3$).
In fact, one can consider a different way to generate the gauge messenger mass,
which would still give $\frac{F^{M_\Phi}}{M_\Phi}\sim \frac{F^C}{C}$.
For instance, the gauge messengers may get a mass through the K\"ahler potential
operator \cite{Nelson:2002sa}
\bea
\label{GM-mass}
\int d^4\theta CC^*\Big(c_\Phi \Phi\Phi^c +{\rm h.c.}\Big),
\eea
where $c_\Phi$ is a generic function of moduli.
In this case, we have
\bea
\frac{F^{M_\Phi}}{M_\Phi}\simeq -2\frac{F^C}{C} \quad \mbox{with}
\quad M_\Phi={\cal O}(m_{3/2}).
\eea
One may consider a more involved model \cite{Hsieh:2006ig}, in which the gauge
messengers get a mass from
\bea
\label{10TeV-DAM}
\int d^4\theta CC^*\Big(\frac{1}{2}{\cal Y}_XX^*X
+ c_\Phi \Phi\Phi^c+\frac{1}{2}c_X X^2\Big)+  \int d^2\theta
C^3\Big(\frac{1}{6}\kappa X^3 +\lambda_\Phi X\Phi\Phi^c\Big) +{\rm h.c.},
\eea
where ${\cal Y}_X,c_\Phi, c_X, \kappa,$ and $\lambda_\Phi$ are again generic
functions of moduli\footnote{Here we assume that all of these coefficients have
real vacuum values.
Unless, the model generically suffers from the SUSY CP problem.}.
One then finds \cite{Hsieh:2006ig}
\bea
\frac{F^{M_\Phi}}{M_\Phi}\simeq
-\frac{8-x_1x_2}{4(1-x_1)}\frac{F^C}{C} \quad \mbox{with}\quad
M_\Phi={\cal O}(m_{3/2}), \eea where \bea
x_1&=&\frac{\lambda_\Phi(3c_X+\sqrt{c_X(c_X-8{\cal Y}_X)}}{2\kappa c_\Phi},
\nonumber \\
x_2&=&\frac{c_X+4{\cal Y}_X-\sqrt{c_X(c_X-8{\cal Y}_X)}}{{\cal Y}_X}.
\eea
It is also possible to have a model in which the ratio
$\frac{F^{M_\Phi}}{M_\Phi}/\frac{F^C}{C}$ takes a positive value of order unity,
while the messenger scale is at an arbitrary intermediate scale
\cite{Okada:2002mv}.
One such an example would be the model with a composite $X$ having an
Affleck-Dine-Seiberg superpotential:
\bea
\label{ADS-potential}
\int d^2\theta C^3 \Big(
\frac{\Lambda_X^{3-l}}{X^l}+\lambda_\Phi X\Phi\Phi^c\Big), \quad
\eea
where $\Lambda_X$ is a dynamical scale hierarchically lower than $M_{GUT}$ and
$l$ is a positive rational number.
One then finds
\bea
\frac{F^{M_\Phi}}{M_\Phi}=\frac{2}{l+1}\frac{F^C}{C}\quad
\mbox{with}\quad M_\Phi\sim
\left(\frac{\Lambda_X^{3+l}}{m_{3/2}}\right)^{1/(l+2)}.
\eea

\section{Soft parameters}

In this section, we examine the renormalization group (RG) running of soft
parameters in deflected mirage mediation.
In particular, we derive (approximate) analytic expressions of low energy soft
parameters, expressed in terms of the SUGRA model parameters defined in the
previous section.
Our results can be used to interpret the TeV scale sparticle masses measured
in future collider experiments within the framework of the most general mixed
mediation scheme preserving flavor and CP.

\subsection{Soft parameters at scales above the gauge threshold scale}

We first examine the soft parameters at scales above the gauge threshold scale
set by the gauge messenger mass $M_\Phi$.
Our starting point is the effective SUGRA action (\ref{effective1}) which has
been obtained after integrating out the flux-stabilized heavy moduli and the
sequestered SUSY breaking sector.
At high scales above $M_\Phi$, but below the gauge coupling unification scale
$M_{GUT}\simeq 2\times 10^{16}\,{\rm GeV}$, the running gauge coupling and the running
K\"ahler metric of the MSSM matter superfield $Q_i$ are given by
\bea
\label{running_above}
\frac{1}{g_a^2(\mu/\sqrt{CC^*})} &=&
{\rm Re}(\tilde{f}_a)-\frac{b_a^H}{16\pi^2}\ln
\left(\frac{\mu^2}{CC^*M_{GUT}^2}\right),
\nonumber \\
\ln {\cal Z}_i(\mu/\sqrt{CC^*})&=&\ln \tilde{\cal Z}_i + \frac{1}{8\pi^2}
\int_{M_{GUT}}^{\mu/\sqrt{CC^*}}\frac{d\mu^\prime}{\mu^\prime}
\gamma_i(\mu^\prime),
\eea
where
\bea
\tilde{f}_a &=& \sum_I k_IT_I+\epsilon,
\nonumber \\
\tilde{\cal Z}_i&=&\prod_I(T_I+T_I^*)^{n_I^i},
\eea
and $b_a^H$ and $\gamma_i$ are the one loop beta function coefficients and
the anomalous dimensions at scales between $M_\Phi$ and $M_{GUT}$:
\bea
b_a^H &=& -3T_a({\rm Adj})+\sum_i T_a(Q_i)
+ \sum_\Phi\left( T_a(\Phi)+T_a(\Phi^c)\right),
\nonumber \\
\gamma_i &=&
2\sum_a C_2^a(Q_i)g_a^2-\frac{1}{2}\sum_{jk}|y_{ijk}|^2,
\eea
where $y_{ijk}$ are the canonical Yukawa couplings given by
\bea
y_{ijk}(\mu) &=&
\frac{\lambda_{ijk}}{\sqrt{e^{-K_0} {\cal Z}_i {\cal Z}_j {\cal Z}_k}}.
\eea
Here we have ignored the $T_I$-dependent K\"ahler and Konishi anomaly
contributions to the running gauge coupling constants
\cite{Kaplunovsky:1994fg}, which are determined by $K_0$ and ${\cal Z}_A$,
and also the UV sensitive string and KK threshold corrections.
Those $T_I$-dependent loop corrections give a contribution of
${\cal O}\left(\frac{F^{T_I}}{8\pi^2}\right)$ to soft parameters, which are
subleading compared to the contributions which will be discussed below.
We also put the superscript $H$ for the high scale beta function coefficients
$b_a^H$ in order to distinguish them from the low scale MSSM beta function
coefficients.
Note that the vacuum value of ${\rm Re}(\tilde{f}_a)$ corresponds to the
unified gauge coupling constant at $M_{GUT}$:
\bea
{\rm Re}(\tilde{f_a})=\sum_Ik_I{\rm Re}(T_I)+{\rm Re}(\epsilon)
= \frac{1}{g_{GUT}^2}.
\eea

The soft SUSY breaking terms are parameterized as
\bea
-{\cal L}_{\rm soft} &=& m^2_i |\phi_i|^2 + \left[ \frac{1}{2} M_a
\lambda^a \lambda^a + \frac{1}{6} A_{ijk} y_{ijk}
\tilde Q_i \tilde Q_j \tilde Q_k +{\rm h.c.}\right],
\eea
where $\lambda^a$ and $\tilde Q_i$ are canonically normalized gauginos and
sfermions, respectively.
Then at scales between $M_\Phi$ and $M_{GUT}$, the running soft parameters
are given by
\bea
\label{running-soft-parameters}
M_a(\mu) &=& -\Big(F^{T_I}\partial_{T_I}+F^C\partial_C\Big)\ln (g_a^2)
\nonumber \\
&=& -F^{T_I}\partial_{T_I}\ln (g_a^2)
+ \frac{b_a^H}{16\pi^2}g_a^2(\mu)\frac{F^C}{C},
\nonumber \\
A_{ijk}(\mu) &=&
-\Big(F^{T_I}\partial_{T_I}+F^C\partial_C\Big) \ln
\left(\frac{\lambda_{ijk}}{e^{-K_0}{\cal Z}_i{\cal Z}_j{\cal Z}_k}\right)
\nonumber \\
&=& F^{T_I}\partial_{T_I}\ln(e^{-K_0}{\cal Z}_i{\cal Z}_j{\cal Z}_k)
- \frac{1}{16\pi^2}(\gamma_i+\gamma_j+\gamma_k)\frac{F^C}{C},
\nonumber \\
m_i^2(\mu)&=&-\Big(F^{T_I}\partial_{T_I}+F^C\partial_C\Big)
\Big(F^{T_J}\partial_{T_J}+F^C\partial_C\Big)^*\ln(e^{-K_0/3}{\cal Z}_i)
\nonumber \\
&=& -F^{T_I}F^{T_J*}\partial_{T_I}\partial_{T^*_J}\ln
(e^{-K_0/3}{\cal Z}_i)+\frac{1}{16\pi^2}
\left(\tilde{\gamma}_i\left(\frac{F^{C}}{C}\right)^*+{\rm h.c.}\right)
- \frac{1}{32\pi^2}\dot{\gamma}_i\left|\frac{F^C}{C}\right|^2,
\eea
where
\bea
\gamma_i &=& 8\pi^2 \frac{d\ln{\cal Z}_i}{d\ln\mu}=2\sum_a
C_2^a(Q_i) g_a^2 - \frac{1}{2} \sum_{jk} |y_{ijk}|^2,
\nonumber \\
\tilde{\gamma}_i&=&F^{T_I}\partial_{T_I}\gamma_i=2\sum_a
C^a_2(Q_i) F^{T_I}\partial_{T_I}g_a^2 +\frac{1}{2} \sum_{jk}
|y_{ijk}|^2F^{T_I}\partial_{T_I}
\ln(e^{-K_0}{\cal Z}_i {\cal Z}_j{\cal Z}_k),
\nonumber \\
\dot{\gamma}_i&=& \frac{d\gamma_i}{d\ln\mu} =
\frac{1}{4\pi^2}\sum_a C^a_2(Q_i)b^H_a g^4_a
+ \frac{1}{16\pi^2}\sum_{jk} |y_{ijk}|^2 (\gamma_i+\gamma_j+\gamma_k).
\eea

The above running soft parameters correspond to the solution of the following
RG equations\footnote{ It is noted that the gaugino masses and
$A$-parameters are a linear superposition of the solutions for two mediations
and $\tilde \gamma_i(\mu)$ is determined by the solution for moduli mediation
at that scale.
Thus, once we obtain the soft parameters for moduli mediation at an arbitrary
scale, we can reconstruct those of mirage mediation without solving the
RG equation again.}:
\bea
\label{eq:RG1}
\frac{d M_a}{d \ln \mu} &=& \frac{b^H_a}{8\pi^2}g_a^2 M_a,
\nonumber \\
\frac{d A_{ijk}}{d \ln \mu}  &=& -\frac{1}{4\pi^2}\sum_a
\Big[C_2^a(Q_i)+C_2^a(Q_j)+C_2^a(Q_k)\Big]g_a^2M_a
\nonumber \\
&& +\,
\frac{1}{16\pi^2}\sum_{lm}\Big(A_{ilm}|y_{ilm}|^2+A_{jlm}|y_{ilm}|^2
+A_{klm}|y_{ilm}|^2\Big),
\nonumber \\
\frac{dm_i^2}{d \ln \mu}  &=& \frac{1}{16\pi^2}\left[-8 \sum_a
C_2^a(Q_i) g_a^2 |M_a|^2
+ \sum_{jk}\Big(m_i^2+m_j^2+m_k^2+|A_{ijk}|^2\Big) |y_{ijk}|^2 \right]
\nonumber \\
&& +\,
\frac{1}{8\pi^2}g_Y^2Y_i  \left[\sum_i Y_i m_i^2 + \sum_\Phi
\Big(Y_\Phi m_\Phi^2+Y_{\Phi^c}m_{\Phi^c}^2\Big)\right],
\eea
with the boundary condition at the scale just below $M_{GUT}$:
\bea
\label{eq:superposition1}
M_a(M_{GUT}) &=& M_0+\frac{b^H_a}{16\pi^2}g_{GUT}^2\frac{F^C}{C},
\nonumber \\
A_{ijk}({M}_{GUT}) &=& \tilde{A}_{ijk}
-\frac{1}{16\pi^2}\Big(\gamma_i(M_{GUT})+\gamma_j(M_{GUT})
+\gamma_k(M_{GUT})\Big)\frac{F^C}{C},
\nonumber \\
{m^2_i}(M_{GUT}) &=& \tilde{m}_i^{2}
+\frac{1}{16\pi^2}\left[\tilde{\gamma}_i(M_{GUT})
\left(\frac{F^C}{C}\right)^*
+ {\rm h.c.}\right] - \frac{1}{32\pi^2}\dot{\gamma}_i(M_{GUT})
\left|\frac{F^C}{C}\right|^2,
\eea
where $Y_i$, $Y_\Phi$ and $Y_{\Phi^c}$ denote the $U(1)_Y$ charges of
$Q_i$, $\Phi$ and $\Phi^c$, respectively, and
\bea
M_0 &\equiv& F^{T_I}\partial_{T_I}\ln
({\rm Re}(\tilde{f}_a))= \frac{g_{GUT}^2}{2}\sum_I k_IF^{T_I},
\nonumber \\
\tilde{A}_{ijk}&\equiv&F^{T_I}\partial_{T_I}\ln (e^{-K_0}
\tilde{\cal Z}_i\tilde{\cal Z}_j\tilde{\cal Z}_k),
\nonumber \\
\tilde{m}_i^2&\equiv&-F^{T_I}F^{T_J*}\partial_{T_I}
\partial_{T^*_J}\ln (e^{-K_0/3}\tilde {\cal Z}_i).
\eea
In view of (\ref{running_above}), $\tilde{f}_a$ and $\tilde{\cal Z}_i$
correspond to the gauge kinetic function and the matter K\"ahler metric
at $M_{GUT}$, and thus $M_0$, $\tilde{A}_{ijk}$ and $\tilde{m}_i^2$
correspond to the moduli-mediated soft parameters at $M_{GUT}$.
As most of our discussion will be independent of their explicit form,
in the following, we will not use any specific form of the moduli K\"ahler
potential $K_0$ and the matter K\"ahler metric $\tilde{\cal Z}_i$,
but instead treat $\tilde{A}_{ijk}$ and $\tilde{m}_i^2$ as family-independent
free parameters constrained by the $SU(5)$ unification relations.

As was  noticed in \cite{Choi:2005uz}, the RG equations (\ref{eq:RG1}) with
the boundary conditions (\ref{eq:superposition1}) have a useful form of
analytic solution.
For the gaugino masses at $\mu > M_\Phi$, one easily finds
\bea
\label{mirage_gaugino}
M_a(\mu) &=& M_0\left[1 +
\frac{b^H_a}{8\pi^2}g_a^2(\mu)\ln\left(\frac{\mu}{M_{\rm mir}}\right)\right],
\eea
with the running gauge coupling constants:
\bea
\frac{1}{g_a^2(\mu)}=\frac{1}{g_{GUT}^2}-\frac{b_a^H}{8\pi^2}
\ln\left(\frac{\mu}{M_{GUT}}\right),
\eea
and the mirage scale $M_{\rm mir}$ given by
\bea
\label{eq:miragescale}
M_{\rm mir} = M_{GUT} \left(\frac{m_{3/2}}{M_{Pl}}\right)^{\alpha/2},
\eea
where $\alpha$ parameterizes the anomaly to moduli mediation ratio:
\bea
\alpha = \frac{F^C/C}{M_0\ln(M_{Pl}/m_{3/2})}\simeq
\frac{m_{3/2}}{M_0\ln(M_{Pl}/m_{3/2})}.
\eea
For the $A$-parameters and sfermion masses, similar analytic expressions are
available {\it if}
\bea
\label{mirage_condition}
&(i)&\mbox{the involved Yukawa couplings are negligible, or}
\nonumber \\
&(ii)&\frac{\tilde{A}_{ijk}}{M_0}
=\frac{\tilde{m}_i^2+\tilde{m}_j^2+\tilde{m}_k^2}{M_0^2}=1\,\,
\mbox{for non-negligible Yukawa coupling $y_{ijk}$}.
\eea
In such cases, one finds \cite{Choi:2005uz}
\bea
\label{eq:mirage1}
A_{ijk}(\mu) &=& \tilde{A}_{ijk}
-\frac{1}{8\pi^2}\Big(\gamma_i(\mu)+\gamma_j(\mu)+\gamma_k(\mu)\Big)M_0
\ln\left(\frac{\mu}{M_{\rm mir}}\right),
\nonumber \\
{m_i^2}(\mu)&=& \tilde{m}_i^2 -\frac{1}{4\pi^2}\gamma_i(\mu)
M_0^2\ln\left(\frac{\mu}{M_{\rm mir}}\right) -\frac{1}{8\pi^2}
\dot{\gamma}_i(\mu) M_0^2\left[\ln\left(\frac{\mu}{M_{\rm
mir}}\right)\right]^2
\nonumber \\
&& +\,\frac{1}{8\pi^2}Y_i {\rm Tr}(Y\tilde{m}^2)
g_Y^2(\mu)\ln\left(\frac{\mu}{M_{GUT}}\right) ,
\eea where
\bea
{\rm Tr}(Y\tilde{m}^2)=\sum_i Y_i\tilde{m}_i^2 +\sum_\Phi
\Big(Y_\Phi\tilde{m}_\Phi^2+Y_{\Phi^c}\tilde{m}_{\Phi^c}^2\Big)
\eea
for the $U(1)_Y$ charge operator $Y$.

The analytic solutions of (\ref{mirage_gaugino}) and (\ref{eq:mirage1})
show that
\beq
M_a(M_{\rm mir}) = M_0, \quad
A_{ijk}(M_{\rm mir}) = \tilde{A}_{ijk},\quad {m_i^2}(M_{\rm mir}) =
\tilde{m}_i^2,
\eeq
if ${\rm Tr}(Y\tilde{m}^2)=0$, which is satisfied for instance when the
moduli-mediated sfermion masses at $M_{GUT}$ satisfy the $SU(5)$ unification
condition and $\tilde{m}_{H_u}^2=\tilde{m}_{H_d}^2$.
In other words, the soft parameters renormalized at $\mu=M_{\rm mir}$ become
identical to the pure moduli-mediated parameters renormalized at $M_{GUT}$,
obeying the unification condition.
With this observation, $M_{\rm mir}$ has been dubbed the mirage messenger
scale as it does not correspond to any physical threshold scale
\cite{Choi:2005uz}.
Note that still the gauge couplings are unified at the conventional GUT scale
$M_{GUT}\simeq 2\times 10^{16}\,{\rm GeV}$.

If there exist non-negligible Yukawa couplings for which the mirage condition
(\ref{mirage_condition}) is not satisfied, the above analytic solutions of
$A_{ijk}$ and $m_i^2$ are not valid anymore.
However, if there is only one such Yukawa coupling, one can still find a
useful analytic expression for the running $A_{ijk}$ and $m_i^2$.
The results are presented in the appendix for $m_i^2$ ($i=H_u, q_3, u_3$) and
$A_{H_uq_3u_3}$ in the MSSM, including only the effects of the top quark
Yukawa coupling.

\subsection{Soft parameters below the gauge threshold scale}

So far, we have discussed the soft parameters at scales above the gauge
threshold scale $M_\Phi$.
Those high scale soft parameters are determined by anomaly and moduli
mediations, and the gaugino and light-family sfermion masses follow the mirage
unification trajectory given by (\ref{mirage_gaugino}) and (\ref{eq:mirage1}).
If there were no exotic matter fields $\Phi+\Phi^c$, soft parameters would
follow these analytic solutions down to the TeV scale.
However, in deflected mirage mediation, soft parameters at scales below
$M_\Phi$ are deflected from the mirage unification trajectory due to the gauge
mediation by $\Phi+\Phi^c$.

Let us examine how the low energy soft parameters are affected by $\Phi+\Phi^c$
which have a mass-superfield:
\bea
\int d^2\theta C^3\Big(M_\Phi+\theta^2 F^{M_\Phi}\Big)\Phi\Phi^c+{\rm h.c.}.
\eea
To compute the low energy soft parameters, one can add the gauge threshold
contribution at $\mu=M_\Phi^-$ to the soft parameters of (\ref{mirage_gaugino})
and (\ref{eq:mirage1}) evaluated at $\mu=M_\Phi^+$, and then apply the RG
equation at lower scales.
(Here $M_\Phi^+$ and $M_\Phi^-$ denote the mass scale just above $M_\Phi$ and
the scale just below $M_\Phi$, respectively.)
The gauge threshold contributions at $M^-_\Phi$ are given by
\bea
\Delta M_a(M_\Phi) &=& -\left.F \cdot
\partial \ln(g_a^2(\mu))\right|_{\mu=M^-_\Phi}
+ \left.F\cdot\partial \ln(g_a^2(\mu))\right|_{\mu=M^+_\Phi}
\nonumber\\
&=& -\frac{N_\Phi}{16\pi^2}\, g_a^2(M_\Phi)
\left(\frac{F^{M_\Phi}}{M_\Phi}+\frac{F^C}{C}\right),
\nonumber \\
\Delta A_{ijk}(M_\Phi) &=& \left.F\cdot \partial \ln ({\cal Z}_i
{\cal Z}_j{\cal Z}_k)\right|_{\mu=M^-_\Phi}
- \left.F\cdot \partial \ln ({\cal Z}_i{\cal Z}_j{\cal Z}_k)
\right|_{\mu=M^+_\Phi}
\nonumber \\
&=& -\frac{1}{16\pi^2}
\Big(\gamma_i(M^-_\Phi) -\gamma_i(M^+_\Phi) \Big)
\left(\frac{F^{M_\Phi}}{M_\Phi}+\frac{F^C}{C}\right) \,=\, 0,
\nonumber \\
\Delta m_i^2(M_\Phi) &=& -\left.(F\cdot\partial)(
\bar{F}\cdot\bar\partial) \ln {\cal Z}_i(\mu)\right|_{\mu=M^-_\Phi}
+\left.(F\cdot\partial)(\bar{F}\cdot\bar{\partial})\ln {\cal Z}_i(\mu)
\right|_{\mu=M^+_\Phi}
\nonumber\\
&=&
-\frac{1}{32\pi^2}\Big(\dot{\gamma}_i(M^-_\Phi)-\dot{\gamma}_i(M^+_\Phi)\Big)
\left|\frac{F^{M_\Phi}}{M_\Phi}+\frac{F^C}{C}\right|^2
\nonumber \\
&=& \frac{N_\Phi}{(16\pi^2)^2}\, 2 C_2^a(\phi^i)\, g_a^4(M_\Phi)\,\left|
\frac{F^{M_\Phi}}{M_\Phi}+\frac{F^C}{C}\right|^2,
\eea
where $F\cdot\partial =F^{T_I}\partial_{T_I}+F^C\partial_C+F^{M_\Phi}
\partial_{M_\Phi}$, and we have assumed that there are $N_\Phi$ flavors of
$\Phi+\Phi^c$ which form the $5+\bar{5}$ representation of $SU(5)$.

We can now obtain the soft parameters  at $\mu<M_\Phi$ by solving the RG
equation with the boundary conditions:
\bea
M_a(M^-_\Phi) &=& M_a(M^+_\Phi)+\Delta M_a(M_\Phi),
\nonumber \\
A_{ijk}(M^-_\Phi)&=&A_{ijk}(M^+_\Phi)+\Delta A_{ijk}(M_\Phi),
\nonumber \\
m_i^2(M^-_\Phi) &=& m_i^2(M^+_\Phi)+\Delta m_i^2(M_\Phi),
\eea
where the soft parameters at $M^+_\Phi$ can be obtained from the high scale
solutions, (\ref{mirage_gaugino}) and (\ref{eq:mirage1}), by replacing $\mu$
with $M^+_\Phi$.
Like the case of high scale solutions, it turns out that the resulting low
energy solutions allow analytic expression which can be used to interpret the
TeV scale sparticle masses.
For instance, gaugino masses are given by \cite{gaugino.code}
\bea
\label{gaugino_def}
M_a(\mu)=M_0^{\rm eff}\left[1+\frac{1}{8\pi^2}b_a g_a^2(\mu)
\ln\left(\frac{\mu}{M_{\rm mir}^{\rm eff}}\right)\,\right],
\eea
where
\bea M_0^{\rm  eff} &=& RM_0,
\nonumber \\
M_{\rm mir}^{\rm eff} &=& M_{GUT}
\left(\frac{m_{3/2}}{M_{Pl}}\right)^{\alpha/2R},
\nonumber \\
b_a&=&-3T_a({\rm Adj})+\sum_i T_a(Q_i)
\eea
for
\bea
M_0 &=& F^{T_I}\partial_{T_I}\ln({\rm Re}(\tilde{f}_a)),
\nonumber \\
R &=& 1 + \frac{N_\Phi g_0^2}{8\pi^2}
\left[\,\frac{\alpha}{2\beta}\ln\left(\frac{M_{Pl}}{m_{3/2}}\right)
- \ln\left(\frac{M_{GUT}}{M_\Phi}\right)\, \right],
\nonumber \\
\alpha &=&
\frac{F^C/C}{M_0\ln(M_{Pl}/m_{3/2})},
\nonumber \\
\beta &=&
-\frac{F^C/C}{F^{M_\Phi}/M_\Phi}.
\eea
Here $\alpha$ and $\beta$ parameterize the anomaly to moduli mediation ratio
and the anomaly to gauge mediation ratio, respectively, $M_0$ is the
moduli-mediated gaugino mass at $M_{GUT}$, and $g^2_0\simeq 1/2$ corresponds
to the unified gauge coupling constant in the absence of $\Phi+\Phi^c$, i.e.
\bea
\frac{1}{g_0^2}=\frac{1}{g_{GUT}^2}+\frac{N_\Phi}{8\pi^2}
\ln\left(\frac{M_{GUT}}{M_\Phi}\right).
\nonumber
\eea

The running $A$-parameters and sfermion masses at $\mu<M_\Phi$ take a more
involved form.
Neglecting the effects of Yukawa couplings, we find
\bea
\label{sfermion_def}
A_{ijk}(\mu) &=& \tilde{A}^{\rm eff}_{ijk}
-\frac{1}{8\pi^2}\Big(\gamma_i(\mu)+\gamma_j(\mu)+\gamma_k(\mu)\Big)
M^{\rm eff}_0\ln\left(\frac{\mu}{M^{\rm eff}_{\rm mir}}\right),
\nonumber \\
{m_i^2}(\mu)&=& \left(\tilde{m}_i^{\rm eff}\right)^2
- \frac{1}{4\pi^2}\gamma_i(\mu) \left(M^{{\rm eff}}_0\right)^2
\ln\left(\frac{\mu}{M^{\rm eff}_{\rm mir}}\right)
- \frac{1}{8\pi^2} \dot{\gamma}_i(\mu) \left(M^{{\rm eff}}_0\right)^2
\left[\ln\left(\frac{\mu}{M^{\rm eff}_{\rm mir}}\right)\right]^2
\nonumber \\
&& +\, \frac{1}{8\pi^2}Y_i \Big(\sum_j Y_jm^2_j(M^-_\Phi) \Big)g^2_Y(\mu)
\ln\left(\frac{\mu}{M_\Phi}\right),
\eea
where
\bea
\label{sfermion_def2}
\tilde A^{\rm eff}_{ijk} &=& \tilde A_{ijk}+
\frac{1}{8\pi^2}(1-R)M_0\Big(\gamma_i(M_\Phi)+\gamma_j(M_\Phi)
+ \gamma_k(M_\Phi)\Big)
\ln\left(\frac{M_{GUT}}{M_\Phi}\right),
\nonumber \\
\left(\tilde m_i^{\rm eff}\right)^2 &=&
\tilde m_i^2 + 2(1-R)^2M_0^2\sum_a
C_2^a(Q_i)\left[\frac{1}{N_\Phi}\frac{g_a^4(M_\Phi)}{g_0^4} \right.
\nonumber \\
&&
\left. +\,\frac{g_a^2(M_\Phi)}{8\pi^2}\left(\frac{1+R}{1-R}
-\frac{g_a^2(M_\Phi)}{g_0^2}\right)
\ln\left(\frac{M_{GUT}}{M_\Phi}\right)\right].
\eea
Here
\bea
\frac{1}{g^2_a(M_\Phi)} &=& \frac{1}{g^2_{GUT}}
+ \frac{b_a+N_\Phi}{8\pi^2} \ln\left(\frac{M_{GUT}}{M_\Phi}\right)
= \frac{1}{g^2_0}+ \frac{b_a}{8\pi^2}
\ln\left(\frac{M_{GUT}}{M_\Phi}\right),
\eea
where $b_a$ are the MSSM beta function coefficients.
For the RG contribution associated with ${\rm Tr}(Y m^2)$, using the
gauge invariance of Yukawa interactions and the anomaly cancellation
conditions, we find
\bea
\sum_i Y_i m^2_i(M^-_\Phi) &=&
\frac{5}{3}\frac{g^2_Y(M_\Phi)}{g^2_0}\sum_i Y_i \tilde m^2_i
+ \left(\frac{5}{3}\frac{g^2_Y(M_\Phi)}{g^2_0} -1\right)
\sum_\Phi(Y_\Phi\tilde{m}_\Phi^2+Y_{\Phi^c}\tilde{m}^2_{\Phi^c}),
\eea
which vanishes when the moduli-mediated sfermion masses at $M_{\rm GUT}$ are
$SU(5)$-invariant and $\tilde{m}^2_{H_u}=\tilde{m}_{H_d}^2$\footnote{
We note that $\tilde A^{\rm eff}_{ijk}$ and $(\tilde m^{\rm eff}_i)^2$ correspond
to the soft parameters at $M^{\rm eff}_{\rm mir}$ after removing the trace term
proportional to $Y_i$.
They are obtained by extrapolating the weak scale soft terms (subtracted
anomaly mediation and the trace term) to $M_{GUT}$ neglecting the gauge
threshold scale.
It is obvious that the low energy soft parameters are summarized in the mirage
mediation pattern using these effective parameters because the superposition
of anomaly mediation and other mediations closes at each scale as in
(\ref{running-soft-parameters}) and anomaly mediation can not distinguish the
origin of other contributions at higher scale \cite{JPS}.
}.

Like the case that soft masses are dominated by one particular mediation, one
can consider the sum rules of sfermion masses in deflected mirage mediation,
which may be useful for identifying the structure of moduli mediation at
$M_{GUT}$.
For instance, from (\ref{sfermion_def}) and (\ref{sfermion_def2}), we find the
following relations amongst the light-family squark and slepton
masses\footnote{After the electroweak symmetry breaking, these sum rules are
affected by the $D$-term contribution, which is of order $M^2_Z$.}:
\bea
\label{sum-rule-1}
&& m^2_{\tilde q_L}(\mu) - 2
m^2_{\tilde u_R}(\mu) + m^2_{\tilde d_R}(\mu) - m^2_{\tilde l_L}(\mu)
+ m^2_{\tilde e_R}(\mu)
\nonumber \\
&& \quad =\, \tilde m^2_{\tilde q_L} - 2 \tilde m^2_{\tilde u_R} + \tilde
m^2_{\tilde d_R} - \tilde m^2_{\tilde l_L} + \tilde
m^2_{\tilde e_R}+\frac{5}{3}\frac{g_Y^2(\mu)}{4\pi^2}
\ln\left(\frac{\mu}{M_\Phi}\right)\sum_iY_im_i^2(M^-_\Phi),
\nonumber \\
&& 2 m^2_{\tilde q_L}(\mu) - m^2_{\tilde u_R}(\mu) - m^2_{\tilde d_R}(\mu)
- 2 m^2_{\tilde l_L}(\mu) + m^2_{\tilde e_R}(\mu)
\nonumber \\
&& \quad =\, 2\tilde m^2_{\tilde q_L} -  \tilde m^2_{\tilde u_R} - \tilde
m^2_{\tilde d_R} - 2\tilde m^2_{\tilde l_L} + \tilde m^2_{\tilde e_R}
+ \frac{4}{3}\frac{g_Y^2(\mu)}{4\pi^2}\ln\left(\frac{\mu}{M_\Phi}\right)
\sum_iY_im_i^2(M^-_\Phi),
\eea
where $\tilde{q}_L$, $\tilde{q}_R=(\tilde{u}_R,\tilde{d}_R)$, $\tilde{l}_L$,
and $\tilde{e}_R$ denote the squark-doublet, squark-singlet, slepton-doublet,
and slepton-singlet, respectively.
If the moduli-mediated sfermion masses at $M_{GUT}$ are $SU(5)$-invariant, i.e.
\bea
\tilde{m}_{\tilde q_L}^2 &=& \tilde{m}_{\tilde u_R}^2\,=\,
\tilde{m}_{\tilde e_R}^2\,=\,\tilde{m}_{10}^2,
\nonumber \\
\tilde{m}_{\tilde d_R}^2&=&\tilde{m}_{\tilde l_L}^2\,=\,\tilde{m}_5^2,
\eea
and also $\tilde{m}_{H_u}^2=\tilde{m}_{H_d}^2$, the above sum rules give
\bea
\label{sum-rules}
&& m^2_{\tilde q_L}(\mu) - 2 m^2_{\tilde u_R}(\mu)
+ m^2_{\tilde d_R}(\mu) - m^2_{\tilde l_L}(\mu) + m^2_{\tilde e_R}(\mu)=0,
\nonumber \\
&& 2 m^2_{\tilde q_L}(\mu) - m^2_{\tilde u_R}(\mu) - m^2_{\tilde d_R}(\mu)
- 2 m^2_{\tilde l_L}(\mu) + m^2_{\tilde e_R}(\mu)=2\tilde{m}_{10}^2
- 3\tilde{m}_5^2,
\eea
indicating that these sum rules can be used to ascertain the existence of
nonzero moduli-mediation as well as the GUT relations of the moduli-mediated
sfermion masses \cite{mass.sum.rule}.

For the effective SUGRA model (\ref{effective1}), which is a representative
class of model for deflected mirage mediation, it is straightforward to
compute $\alpha$ and $\beta$, which gives
\bea
\alpha &\simeq& 1+ \frac{{\rm Re}(\epsilon)}{\sum_I k_I{\rm Re}(T_I)},
\nonumber \\
\beta &=& \frac{n-1}{2} \qquad (n\geq 3),
\eea
where we have used $\ln(A_I/w_0)\simeq \ln(M_{Pl}/m_{3/2})$.
Here, the value of $\beta$ for $n>3$ applies to the model in which $X$ is
stabilized by the non-renormalizable superpotential term
$\kappa X^n/M_{Pl}^{n-3}$, while the value of $\beta$ for $n=3$ applies to
the model with $\kappa=0$, in which $X$ is stabilized by the radiative
correction to its K\"ahler potential.
In view of underlying string theory, ${\rm Re}(\epsilon)$ corresponds to a
higher order correction to the gauge kinetic function in the $g_{st}$ or
$\alpha^\prime$ expansion. This suggests that ${\rm Re}(\epsilon)$ is
significantly smaller than $\sum_I k_I{\rm Re}(T_I)$, and thus $\alpha$ has
a value close to the unity.

\begin{figure}[h]
\begin{center}
\includegraphics[width=0.45\textwidth, clip]{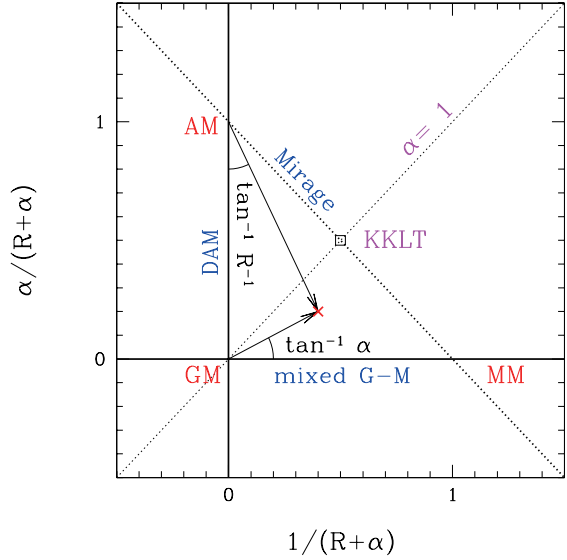}
\caption{Parameter space of deflected mirage mediation spanned by
$\alpha$ and $R$. Here AM, GM and MM denote anomaly mediation, gauge
mediation and moduli mediation, respectively.} \label{fig:triangle}
\end{center}
\end{figure}

With (\ref{gaugino_def}) and (\ref{sfermion_def}), providing the analytic
expression of low energy soft parameters in deflected mirage mediation, one
can take an appropriate limit to obtain the soft parameters in more familiar
case dominated by a single mediation.
Specifically, each single mediation corresponds to the limit:
\bea
&*& \makebox[3.8cm][l]{Anomaly mediation} :\,  R\rightarrow 1, \quad
\frac{1}{\alpha}\rightarrow 0, \quad \alpha M_0 = \mbox{finite},
\quad \tilde{A}_{ijk}=\tilde{m}_i^2=0,
\nonumber \\
&*& \makebox[3.8cm][l]{Gauge mediation} :\,  \frac{1}{R}\rightarrow 0, \quad
RM_0=\mbox{finite}, \quad \tilde{A}_{ijk}=\tilde{m}_i^2=0,
\nonumber \\
&*& \makebox[3.8cm][l]{Moduli mediation} :\, R\rightarrow 1, \quad
\alpha\rightarrow 0, \nonumber
\eea
while the mixed gauge-anomaly mediation ($=$ deflected anomaly mediation) and
the mixed moduli-anomaly mediation ($=$ mirage mediation) can be obtained as
\bea
&*& \makebox[5.4cm][l]{Deflected anomaly mediation} :\,\frac{1}{R}\rightarrow 0,
\quad \frac{\alpha}{R}=\mbox{finite}, \nonumber \\
&& \makebox[5.85cm][l]{}
RM_0=\mbox{finite},
\quad \tilde{A}_{ijk}=\tilde{m}_i^2=0,
\nonumber \\
&*& \makebox[5.4cm][l]{Mirage mediation}:\, R\rightarrow 1,
\nonumber \eea
Fig$.$ \ref{fig:triangle} summarizes these different limits of deflected mirage
mediation in the parameter space spanned by $\alpha$ and $R$.

Soft parameters in case of multi-step gauge thresholds can be obtained by
applying our results recursively.
For instance, the gaugino masses after the $n$-step of  thresholds are given by
(\ref{gaugino_def}) with
\bea
M^{\rm eff}_0 &=& R_n R_{n-1} \cdots R_1 M_0,
\nonumber \\
\alpha_{\rm eff} &=& \frac{\alpha}{R_n R_{n-1} \cdots R_1},
\eea
where
\bea
R_n &=& 1 + \frac{N_{\Phi_n} g^2_0}{8\pi^2}
\left[ \frac{1}{R_{n-1} \cdots R_1}
\frac{\alpha}{2\beta_n}\ln\left(\frac{M_{Pl}}{m_{3/2}}\right) -
\ln\left(\frac{M_{GUT}}{M_{\Phi_n}}\right)\right]
\nonumber
\eea
for $N_{\Phi_n}$ denoting the number of the gauge messenger pairs at the $n$-th
threshold scale $M_{\Phi_n}$, and $\beta_n$ is the anomaly to gauge mediation
ratio for the $n$-th gauge threshold. Light-family sfermion soft parameters also
can be written as (\ref{sfermion_def}) with appropriately defined
$\tilde{A}_{ijk}^{\rm eff}$ and $\tilde{m}_i^{\rm eff}$.
As we will see below, such parametrization provides a useful set-up to interpret
the TeV scale sparticle masses within the framework of the most general mixed
moduli-anomaly-gauge mediation.

\subsection{Sparticle masses at the TeV scale}

From (\ref{gaugino_def}) and (\ref{sfermion_def}), one can obtain the low energy
sparticle masses at the TeV scale.
If one assumes that the moduli-mediated sfermion masses at $M_{GUT}$ satisfy the
$SU(5)$ unification condition and also $\tilde{m}_{H_u}^2=\tilde{m}_{H_d}^2$,
the gaugino and light-family sfermion masses in generic deflected mirage
mediation at the renormalization point $\mu=500\,{\rm GeV}$ are given by\footnote{
For colored sparticles, there can be a sizable difference between this running mass
at $\mu=500\,{\rm GeV}$ and the physical mass \cite{martin}.}
\bea
\label{DMM-soft-masses}
M_1 &=& M^{\rm eff}_0 \left[ 0.43  + 0.29 \alpha_{\rm eff} \right],
\nonumber \\
M_2 &=& M^{\rm eff}_0 \left[ 0.83  + 0.084 \alpha_{\rm eff}\right],
\nonumber \\
M_3 &=& M^{\rm eff}_0 \left[ 2.5  - 0.74 \alpha_{\rm eff} \right],
\nonumber \\
m^2_{\tilde q_L} &=& \tilde m_{10}^2+ (M^{\rm eff}_0)^2\left[5.0 - 3.48
\alpha_{\rm eff} + 0.48 \alpha_{\rm eff}^2+\delta_{\tilde q_L} \right],
\nonumber \\
m^2_{\tilde u_R} &=& \tilde m_{10}^2 + (M^{\rm eff}_0)^2\left[4.6 - 3.29
\alpha_{\rm eff}   + 0.49\alpha_{\rm eff}^2+\delta_{\tilde u_R} \right],
\nonumber \\
m^2_{\tilde e_R} &=& \tilde m_{10}^2 + (M^{\rm eff}_0)^2\left[0.15 -0.045
\alpha_{\rm eff} - 0.015\alpha_{\rm eff}^2+\delta_{\tilde e_R} \right],
\nonumber \\
m^2_{\tilde d_R} &=& \tilde m_5^2 + (M^{\rm eff}_0)^2\left[4.5 - 3.27
\alpha_{\rm eff}  + 0.49\alpha_{\rm eff}^2 +\delta_{\tilde d_R}\right],
\nonumber \\
m^2_{\tilde l_L} &=& \tilde m_5^2 + (M^{\rm eff}_0)^2\left[0.5 - 0.22
\alpha_{\rm eff} - 0.014\alpha_{\rm eff}^2 +\delta_{\tilde l_L} \right],
\eea
where
\bea
M_0^{\rm eff}&=&R M_0,\quad \alpha_{\rm eff} \,=\,\alpha/R,
\nonumber \\
\delta_i &\equiv& \frac{\left(\tilde{m}_i^{\rm eff}\right)^2
-\tilde{m}_i^2}{\left(M_0^{\rm eff}\right)^2} =\sum_a
C_2^a(Q_i)\delta_a,
\eea
for
\bea
\delta_a &=& \frac{2(1-R)^2}{R^2}
\left[ \frac{1}{N_\Phi}\frac{g^4_a(M_\Phi)}{g^4_0} +
\frac{g^2_a(M_\Phi)}{8\pi^2} \left( \frac{1+R}{1-R} -
\frac{g^2_a(M_\Phi)}{g^2_0} \right)
\ln\left(\frac{M_{GUT}}{M_\Phi}\right) \right].
\eea
One interesting limit of deflected mirage mediation is the pure mirage
mediation in which there is no gauge-mediated contribution.
In this limit, $R=1$, and therefore
\bea
\label{miragepattern}
M_0^{\rm eff}=M_0,\quad\alpha_{\rm eff}=\alpha, \quad \delta_i=0.
\eea

Since the deflected mirage mediation provides a framework that involves all
three prominent flavor and CP conserving mediation mechanisms, it is important
to understand how does each mediation reveal its existence in low energy
sparticle masses.
From (\ref{DMM-soft-masses}), one easily notices that anomaly mediation reveals
itself through a nonzero value of $\alpha_{\rm eff}$, which can be read off
from the gaugino mass pattern \cite{gaugino.code}.
Once $M_0^{\rm eff}$ and $\alpha_{\rm eff}$ could be determined from the
gaugino masses, one may examine $m_{\tilde{q}_L}^2-m_{\tilde{e}_R}^2$ and
$m_{\tilde{d}_R}^2-m_{\tilde{l}_L}^2$ to see the existence of gauge mediation,
from which $\delta_{\tilde{q}_L}-\delta_{\tilde{e}_R}$ and
$\delta_{\tilde{d}_R}-\delta_{\tilde{l}_L}$ can be determined.

It is obvious that  $\delta_i$, particularly
$\delta_{\tilde{q}_L}-\delta_{\tilde{e}_R}$ and
$\delta_{\tilde{d}_R}-\delta_{\tilde{l}_L}$, are crucial for identifying the
underlying mediation mechanism from the sparticle masses at TeV.
Let us thus examine the possible values of $\delta_i$ in various models of
deflected mirage mediation.
The overall size of $\delta_i$ is determined by
\bea
R-1 &=& \frac{N_\Phi g_0^2}{8\pi^2}
\left[\frac{\alpha}{2\beta}\ln\left(\frac{M_{Pl}}{m_{3/2}}\right)
-\ln\left(\frac{M_{GUT}}{M_\Phi}\right)\right],
\eea
where $\alpha/\beta$ represents the gauge to moduli mediation ratio.
If $\alpha/\beta>0$, there is a cancellation between gauge and moduli
mediations, reducing the size of $R-1$, and thus of $\delta_i$.
In particular, if the gauge messenger mass $M_\Phi$ is close to the scale
\bea
\label{nd-scale}
M_{GUT}\left(\frac{m_{3/2}}{M_{Pl}}\right)^{\alpha/2\beta} \simeq
10^{16.3-6.9\alpha/\beta}\,{\rm GeV},
\eea
the cancellation is most efficient.
For many of the representative SUGRA models discussed in the previous section,
although the strength of gauge mediation is comparable to those of anomaly
and moduli mediations, the resulting $\delta_i$ are small because of this
cancellation.
In such models, the predicted pattern of sparticle masses is quite similar to
that of pure mirage mediation.

\begin{figure}[t]
\begin{center}
\begin{minipage}{16cm}
\centerline{
\hspace*{-0.1cm} \epsfig{figure=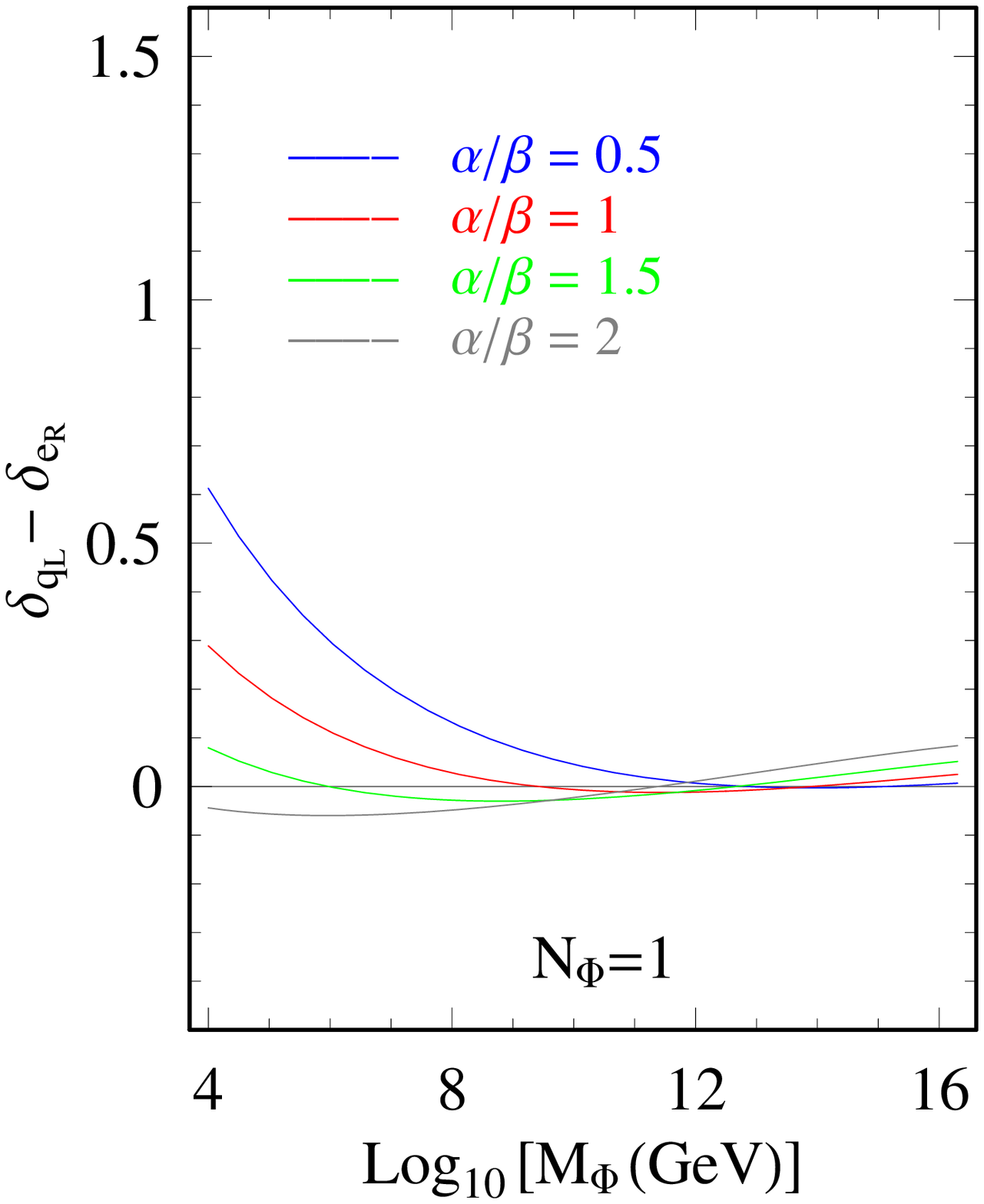,angle=0,width=4.9cm}
\hspace*{0.1cm}  \epsfig{figure=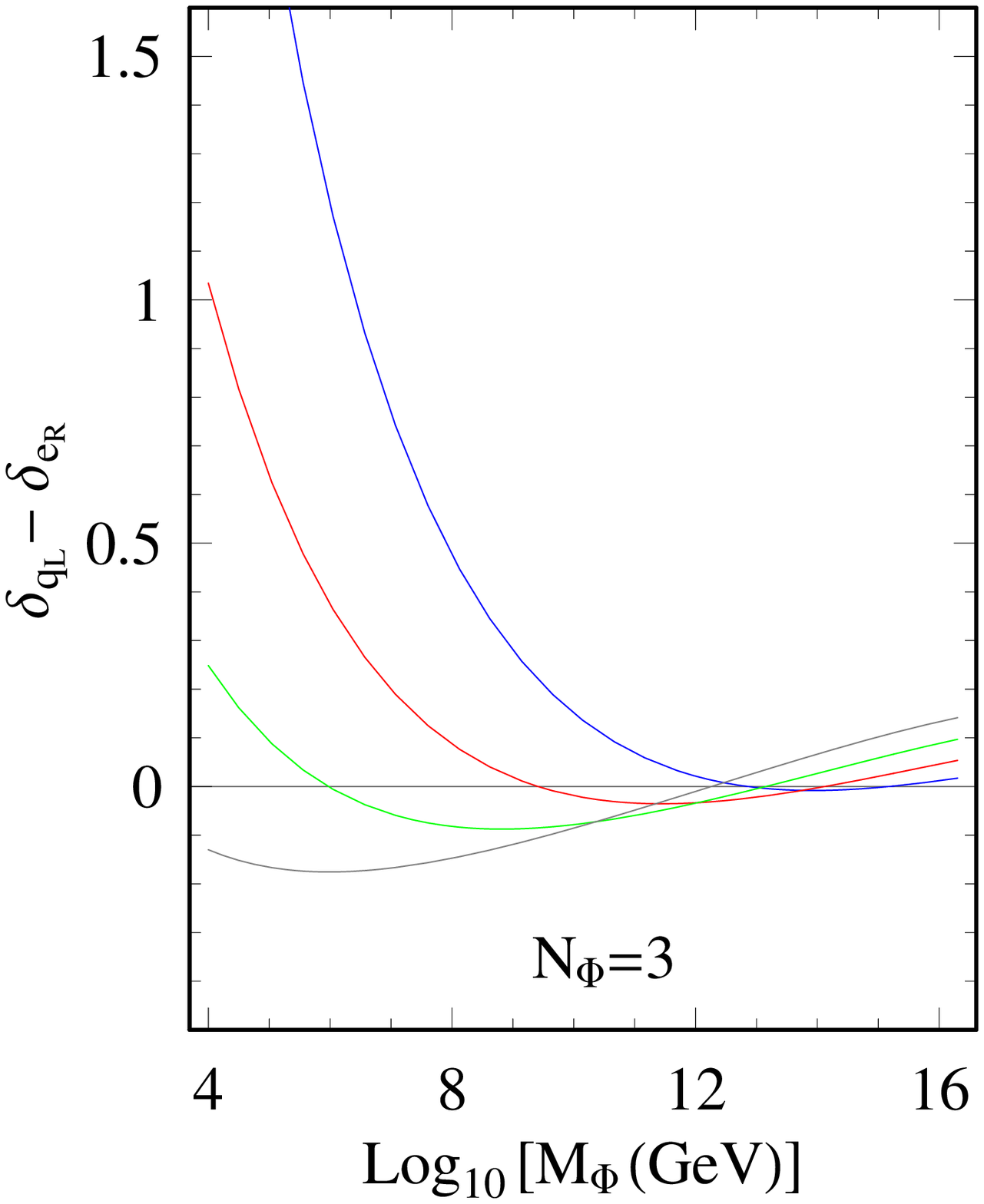,angle=0,width=4.9cm}
\hspace*{0.1cm}  \epsfig{figure=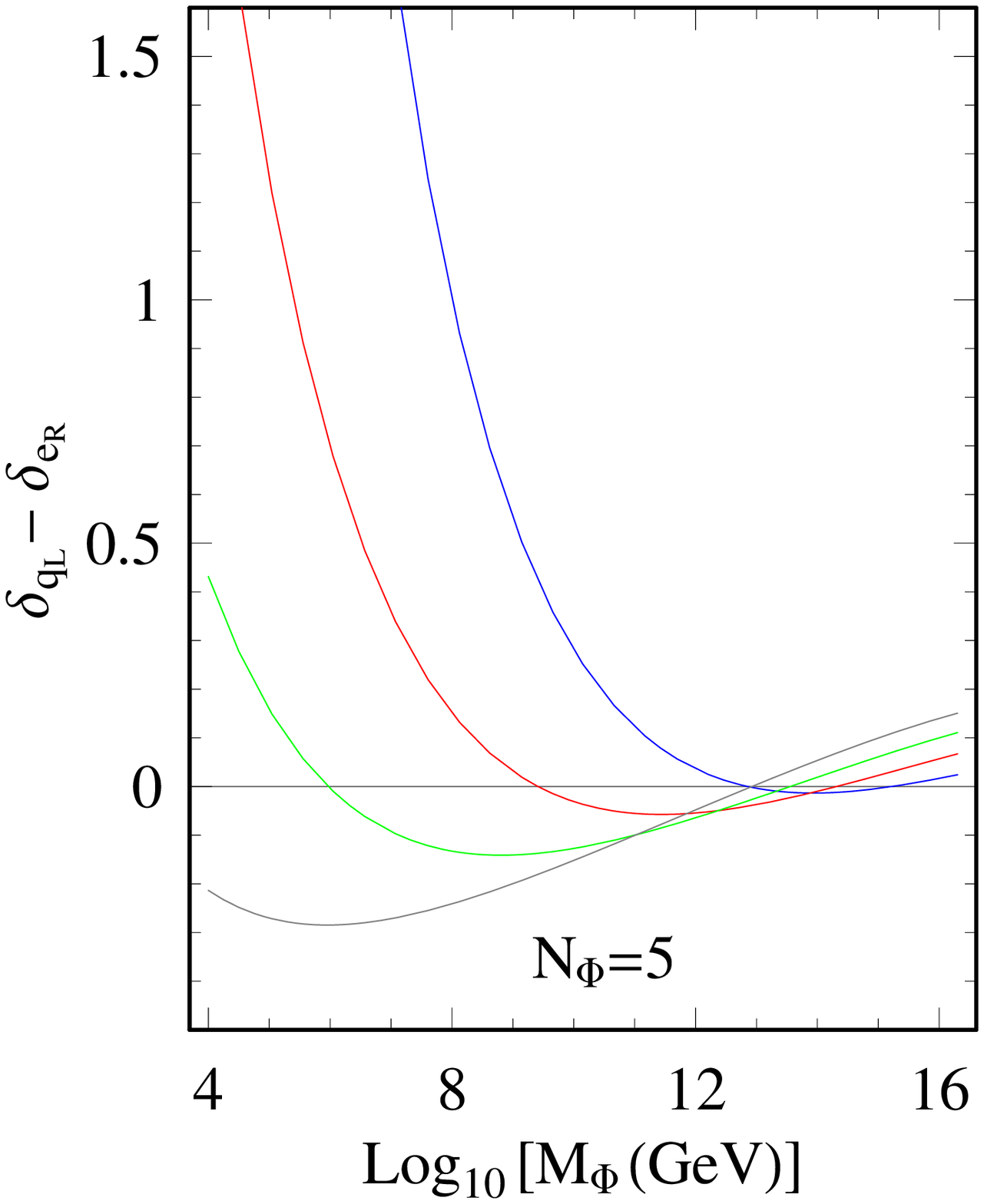,angle=0,width=4.9cm} }
\centerline{
\hspace*{-0.1cm} \epsfig{figure=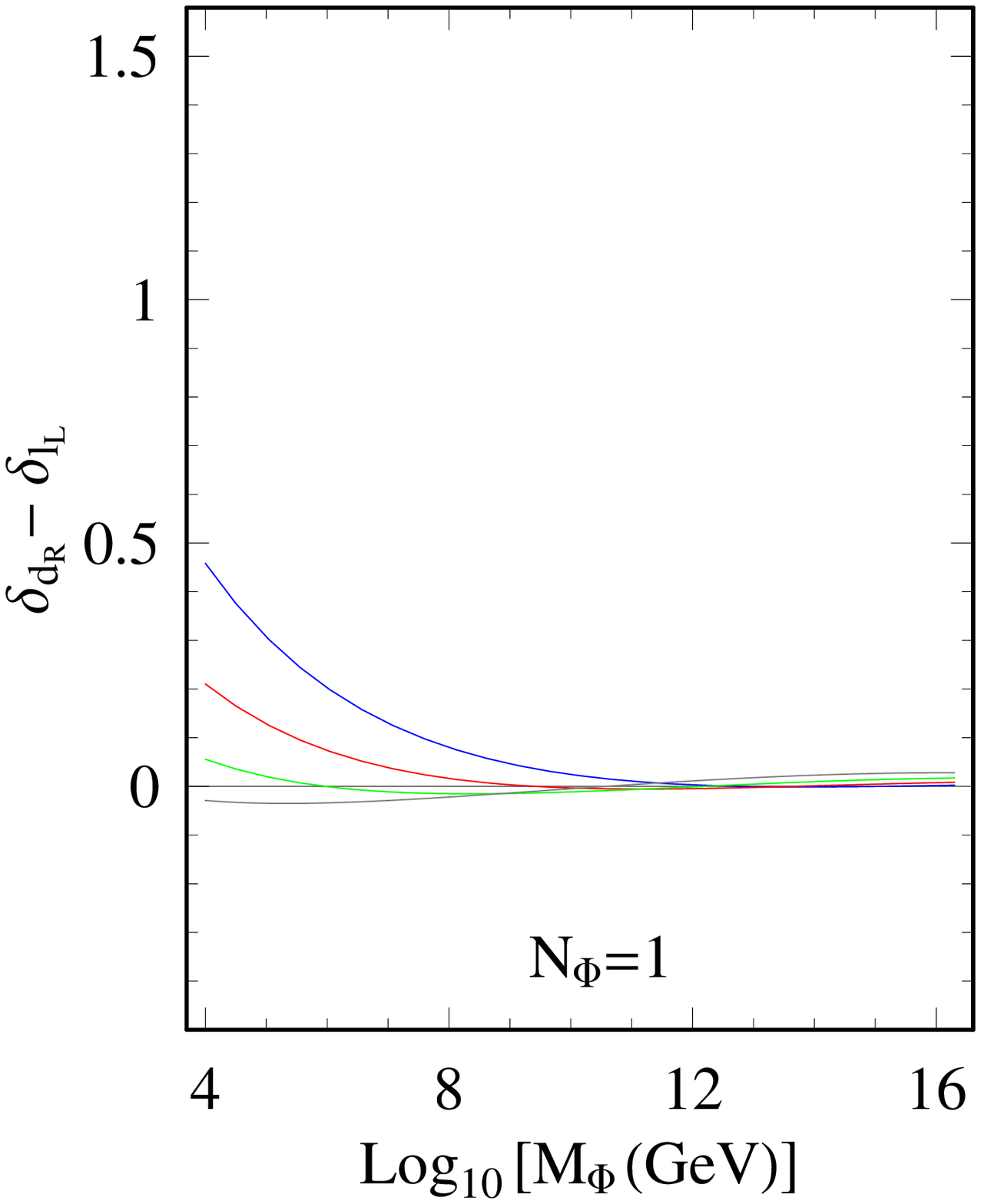,angle=0,width=4.9cm}
\hspace*{0.1cm}  \epsfig{figure=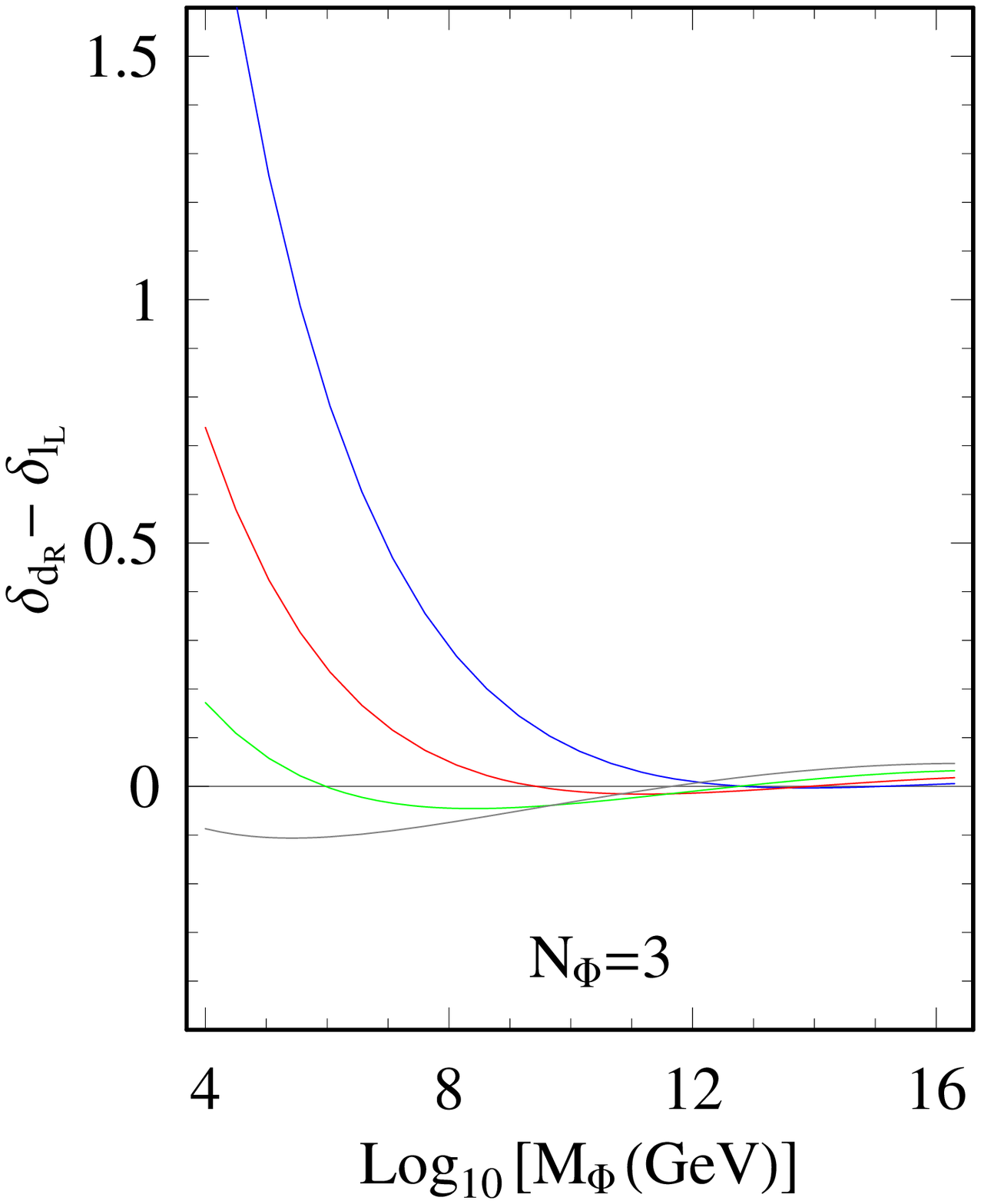,angle=0,width=4.9cm}
\hspace*{0.1cm}  \epsfig{figure=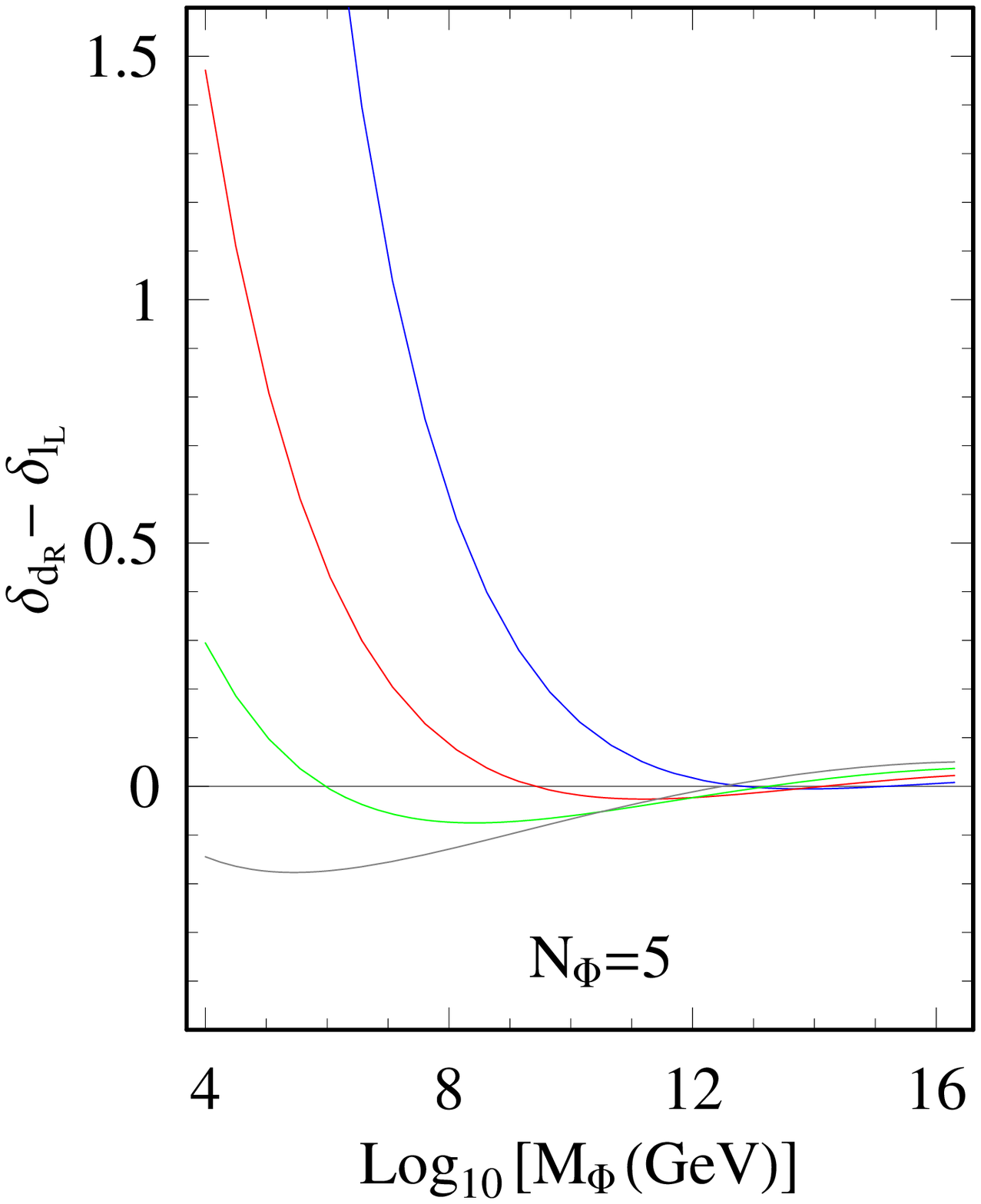,angle=0,width=4.9cm} }
\caption{Difference of $\delta_i$ in deflected mirage mediation with
a generic value of $M_\Phi$.
The upper panels show $\delta_{\tilde q_L}-\delta_{\tilde e_R}$ while
the lower ones show
$\delta_{\tilde d_R}-\delta_{\tilde l_L}$ for $N_\Phi=1,3,5$.
\label{DMM-N}}
\end{minipage}
\end{center}
\end{figure}

Let us first examine  $\delta_{\tilde{q}_L}-\delta_{\tilde{e}_R}$ and
$\delta_{\tilde{d}_R}-\delta_{\tilde{l}_L}$ in the effective SUGRA model
(\ref{effective1}).
In this model with $\kappa\neq 0$, the gauge messenger mass is induced by the
superpotential coupling $\lambda_\Phi X\Phi\Phi^c$ with $X$ stabilized by
$\kappa X^n/M^{n-3}_{Pl}\,(n>3)$, which results in
\bea
\label{higher-dim-operator}
\beta &=& \frac{n-1}{2}, \quad M_{\Phi} \,=\, x
M_{Pl}\left(\frac{m_{3/2}}{M_{Pl}}\right)^{1/(n-2)},
\eea
where $x=\lambda_\Phi \kappa^{1/(2-n)}$.
As $\alpha\simeq 1$ at leading order in the $g_{st}$ or $\alpha^\prime$
expansion in underlying string theory, we first focus on the case with
$\alpha=1$.
We then find
\bea
|\,\delta_{\tilde q_L} - \delta_{\tilde e_R}| < 0.02N_\Phi, \quad
|\,\delta_{\tilde d_R} - \delta_{\tilde l_L}| < 0.01N_\Phi,
\eea
for the parameter range: $\alpha=1$, $4\leq n\leq 6$ and $10^{-(n-3)}\leq x$.
For different value of $\alpha$, they can have a bigger value, but still
bounded as
\bea
|\, \delta_{\tilde q_L} - \delta_{\tilde e_R}| < 0.04N_\Phi, \quad
|\,\delta_{\tilde d_R} - \delta_{\tilde l_L}| < 0.02N_\Phi,
\eea
for the parameter range:
$0.5\leq \alpha\leq 2$, $4\leq n\leq 6$, $10^{-(n-4)}\leq x$,
and $N_\Phi\leq 8$.
A less stringent bound is obtained for the axionic mirage mediation model
\cite{Nakamura:2008ey}, in which $\kappa=0$ and $X$ is stabilized by the
radiative correction to its K\"ahler potential, yielding $\beta=1$.
Provided that $\langle X\rangle$ is fixed at a scale between $10^9\,{\rm GeV}$
and $10^{12}\,{\rm GeV}$ as required for ${\rm Im}(X)$ to be the QCD axion,
it is found that
\bea
|\,\delta_{\tilde q_L} - \delta_{\tilde e_R}| < 0.16N_\Phi, \quad
|\,\delta_{\tilde d_R} - \delta_{\tilde l_L}| < 0.08N_\Phi,
\eea
for $0.5\leq \alpha \leq 2$ and $N_\Phi\leq 8$.
The above results for the models of (\ref{effective1}) show that
$\delta_{\tilde{q}_L}-\delta_{\tilde{e}_R}$ and
$\delta_{\tilde{d}_R}-\delta_{\tilde{l}_L}$ are small over a reasonable range
of model parameters, and therefore the predicted sparticle mass pattern is
close to the pure mirage pattern obtained from (\ref{miragepattern}).
In Fig$.$ \ref{DMM-N}, we depict the values of
$\delta_{\tilde{q}_L}-\delta_{\tilde{e}_R}$ and
$\delta_{\tilde{d}_R}-\delta_{\tilde{l}_L}$ in deflected mirage mediation
scenario with a generic value of $M_\Phi$, where $\alpha,\beta$ and $N_\Phi$
are assumed as $0.5\leq \alpha/\beta\leq 2$ and $1\leq N_\Phi\leq 5$.
The models of (\ref{effective1}) typically give $M_\Phi\geq 10^9\,{\rm GeV}$ and
$0.5\leq \alpha/\beta\leq 2$, for which
$\delta_{\tilde{q}_L}-\delta_{\tilde{e}_R}$ and
$\delta_{\tilde{d}_R}-\delta_{\tilde{l}_L}$ have a small value as long as
$N_\Phi$ is not unreasonably large.

\begin{figure}[t]
\begin{center}
\begin{minipage}{16cm}
\centerline{
\hspace*{-0.2cm} \epsfig{figure=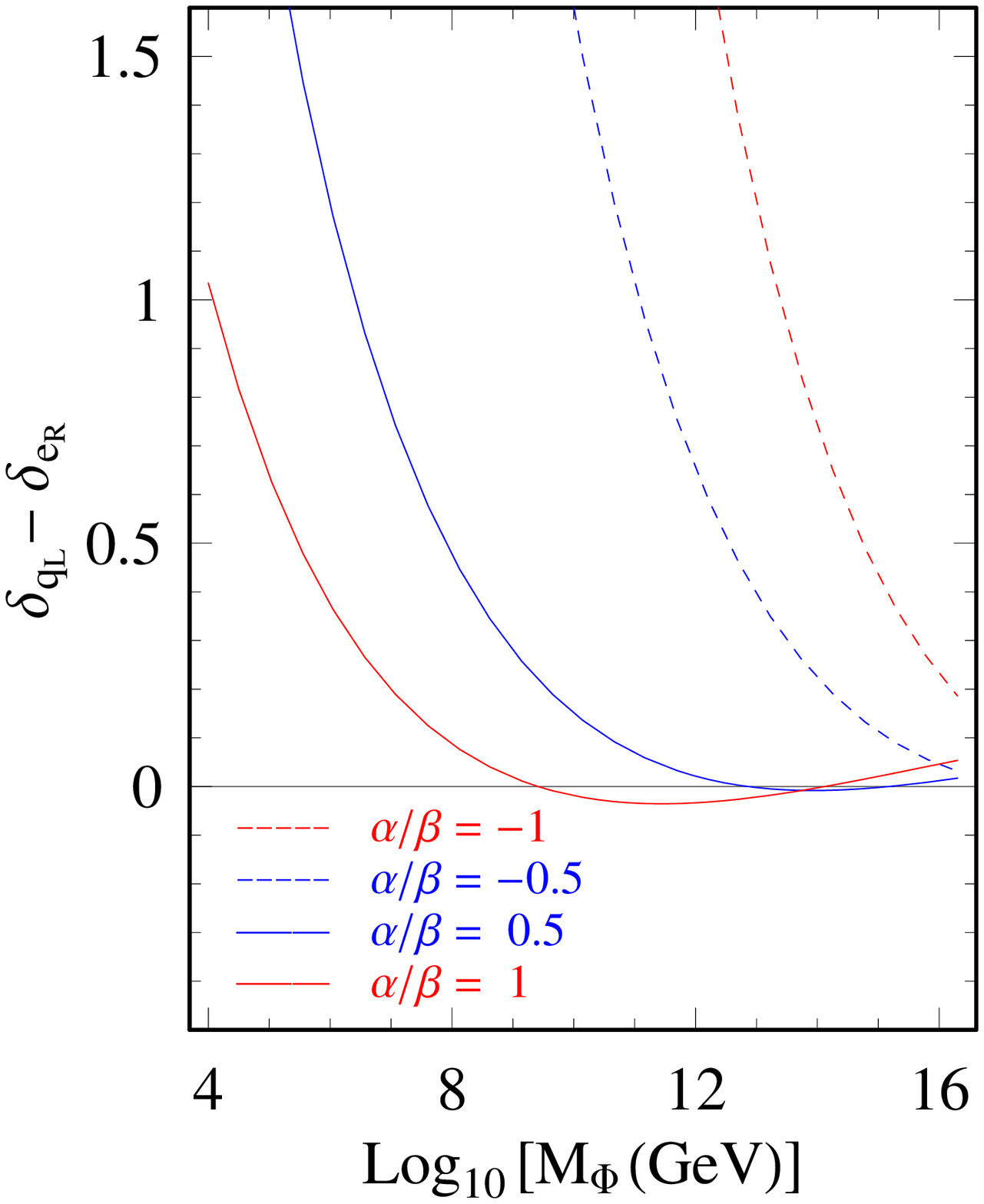,angle=0,width=4.9cm}
\hspace{0.1cm}   \epsfig{figure=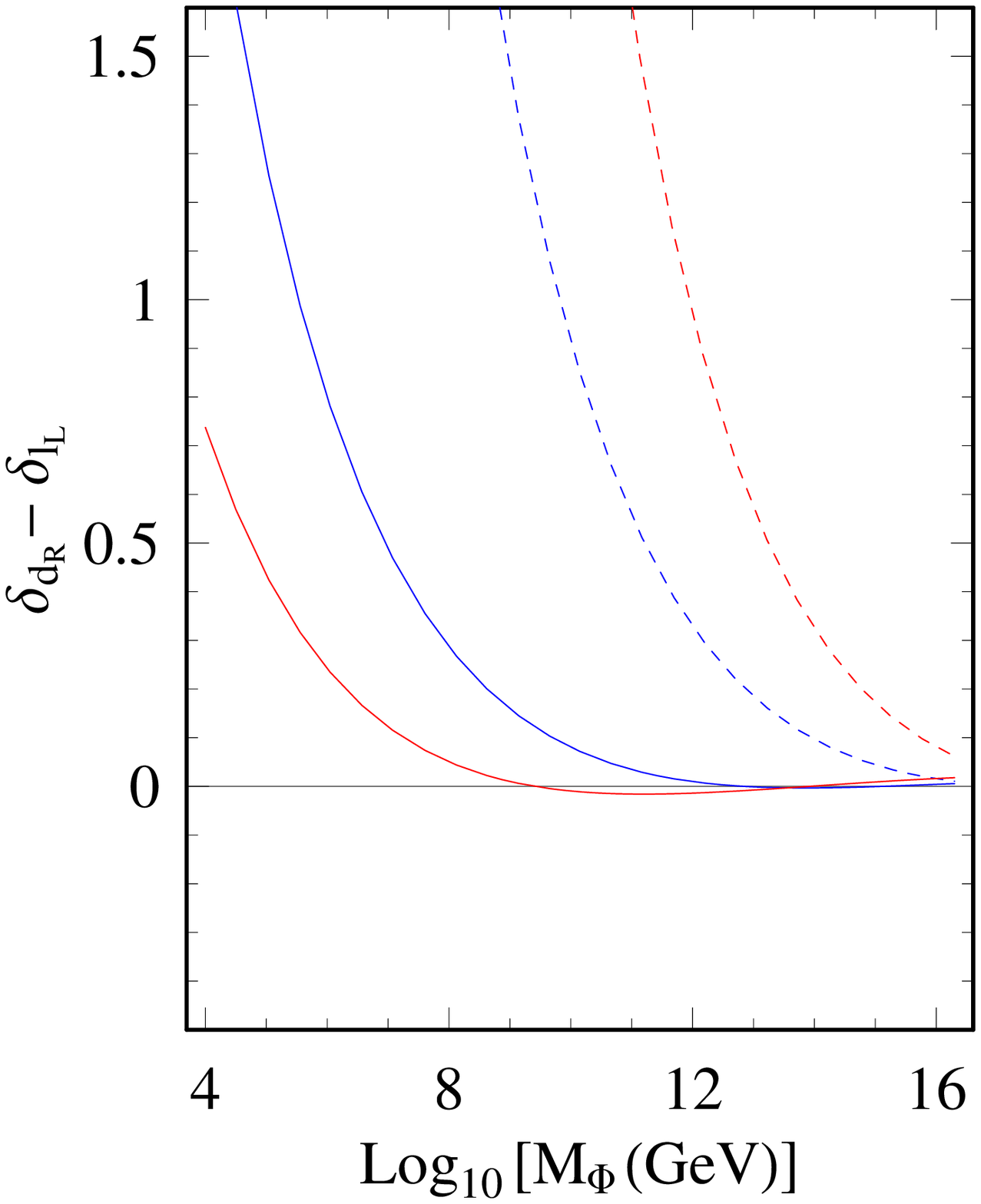,angle=0,width=4.9cm} }
\caption{Difference of $\delta_i$ in deflected mirage mediation.
The solid lines show the values for $0.5\leq \alpha/\beta\leq 1$ while
the dashed ones for $-1\leq \alpha/\beta\leq -0.5$, with $N_\Phi=3$.
\label{DMM-P}}
\end{minipage}
\end{center}
\end{figure}

There are in fact some models which can give a sizable value of
$\delta_{\tilde{q}_L}-\delta_{\tilde{e}_R}$ and
$\delta_{\tilde{d}_R}-\delta_{\tilde{l}_L}$.
One such example is a model with a negative value of $\alpha/\beta$.
For an illustration, we depict in Fig$.$ \ref{DMM-P} the values of
$\delta_{\tilde{q}_L}-\delta_{\tilde{e}_R}$ and
$\delta_{\tilde{d}_R}-\delta_{\tilde{l}_L}$ for
$-1\leq \alpha/\beta\leq -0.5$, and compare them with the values for
$0.5\leq \alpha/\beta\leq 1$.

Another scheme  which can give a sizable value of
$\delta_{\tilde{q}_L}-\delta_{\tilde{e}_R}$ and
$\delta_{\tilde{d}_R}-\delta_{\tilde{l}_L}$ would be the deflected anomaly
mediation \cite{Pomarol:1999ie,Nelson:2002sa,Hsieh:2006ig,Okada:2002mv,Endo:2008gi},
in which there is no moduli mediation. Soft parameters in deflected anomaly
mediation can be obtained by taking the limit:
$1/R\rightarrow 0$ and $\tilde A_{ijk}=\tilde m^2_i=0$,
while keeping $RM_0$ and $\alpha/R$ to have a nonzero finite value.
The resulting sparticle masses at $\mu=500\,{\rm GeV}$ are given by
(\ref{DMM-soft-masses}) with
\bea
M^{\rm eff}_0 &=& \frac{N_\Phi
g^2_0}{16\pi^2}\frac{m_{3/2}}{\beta}\simeq 3\times 10^{-3}
\frac{N_\Phi m_{3/2}}{\beta},
\nonumber \\
\alpha_{\rm eff} &=& \frac{16\pi^2}{g^2_0\ln(M_{Pl}/m_{3/2})}
\frac{\beta}{N_\Phi} \simeq
\frac{10\beta}{N_\Phi},
\nonumber \\
\left(\tilde m^{{\rm eff}}_i\right)^2 &=& \left(M^{{\rm eff}}_0\right)^2
\delta_i, \eea where $\delta_i=\sum_a
C^a_2(Q_i)\delta_a$ with \bea \delta_a &=& 2\left[\frac{1}{N_\Phi}
\frac{g^4_a(M_\Phi)}{g^4_0} - \frac{g^2_a(M_\Phi)}{8\pi^2}\left( 1 +
\frac{g^2_a(M_\Phi)}{g^2_0} \right)
\ln\left(\frac{M_{GUT}}{M_\Phi}\right)\right].
\eea
Thus, in deflected anomaly mediation limit, $\delta_i$ are determined by just
$M_\Phi$ and $N_\Phi$.

\begin{figure}[t]
\begin{center}
\begin{minipage}{16cm}
\centerline{
\hspace*{-0.2cm} \epsfig{figure=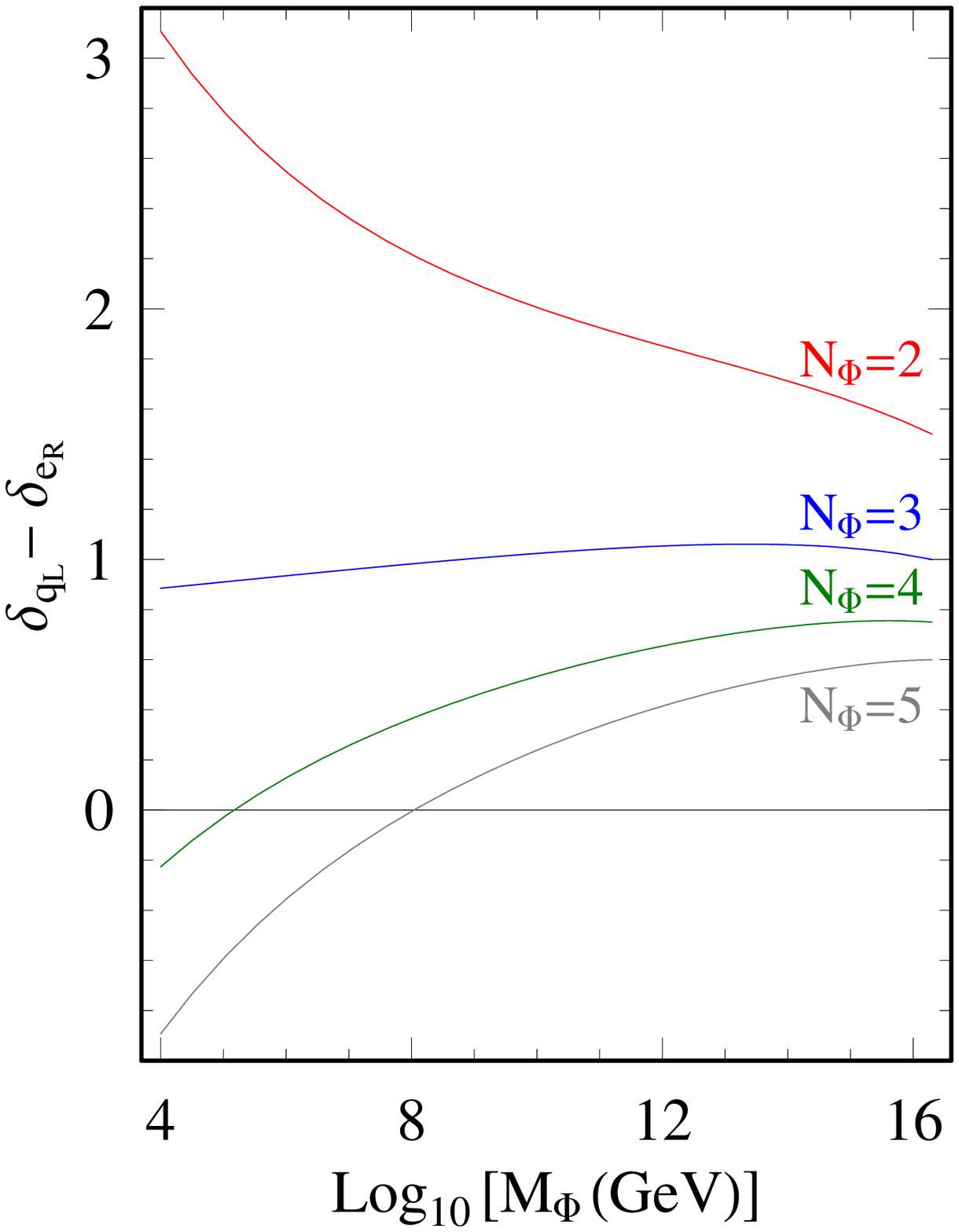,angle=0,width=4.9cm}
\hspace*{0.1cm}  \epsfig{figure=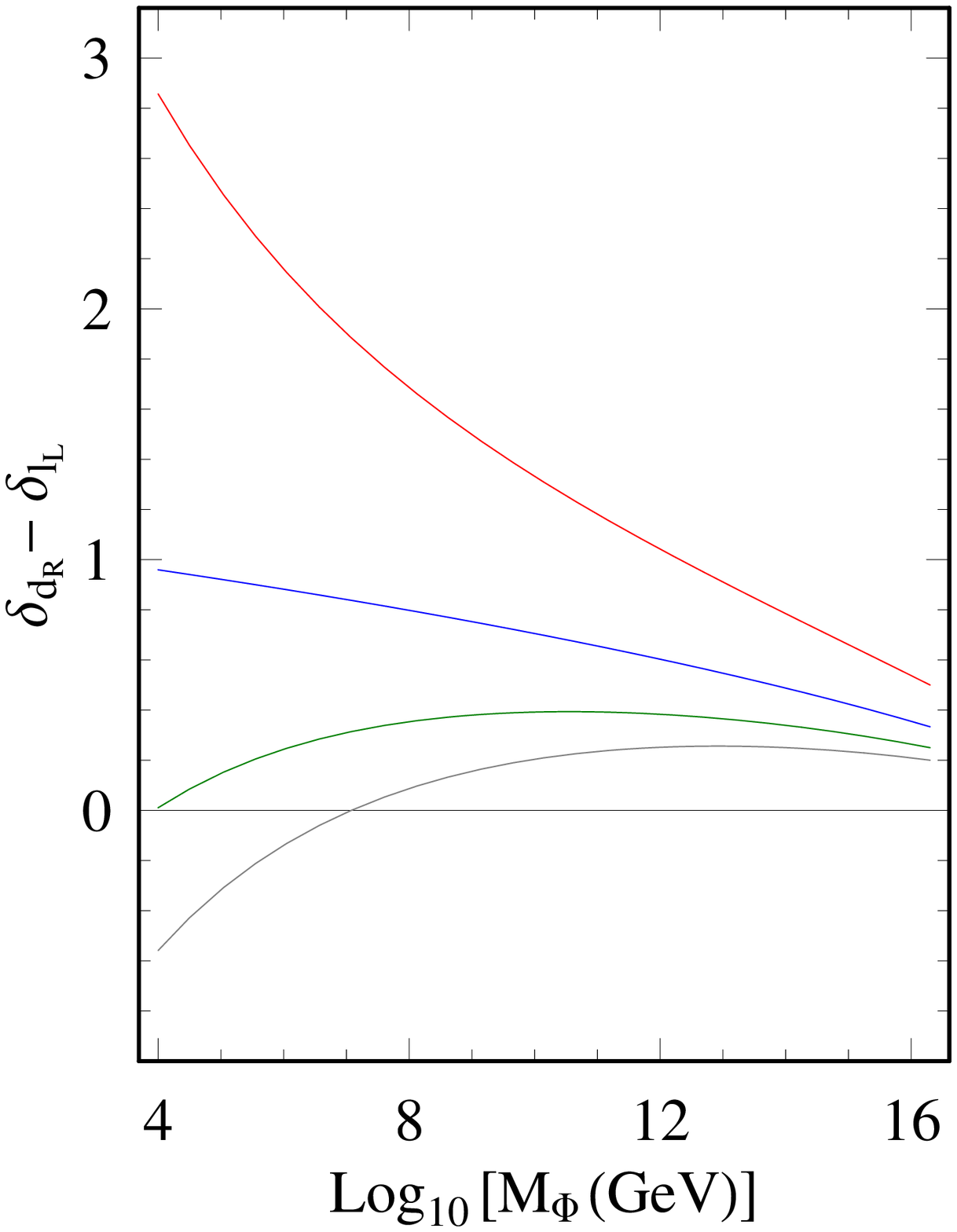,angle=0,width=4.9cm} }
\caption{Difference of $\delta_i$ in deflected anomaly mediation
for $2\leq N_\Phi \leq 5$ with a generic value of $M_\Phi$.
\label{DAM-F}}
\end{minipage}
\end{center}
\end{figure}

Let us examine the values of
$\delta_{\tilde{q}_L}-\delta_{\tilde{e}_R}$ and
$\delta_{\tilde{d}_R}-\delta_{\tilde{l}_L}$ in some specific models of
deflected anomaly mediation.
For the model (\ref{10TeV-DAM}), we have $M_\Phi={\cal O}(m_{3/2})$, while
$\beta$ can take any value of order unity.
Keeping the perturbative gauge coupling unification requires $N_\Phi\leq 5$,
and tachyonic slepton can be avoided for
\bea
-0.25-0.35 N_\Phi \,\lesssim\, \beta \,\lesssim\, 0.25+0.05 N_\Phi.
\eea
We then find
\bea
\label{DAM-1}
\delta_{\tilde q_L}- \delta_{\tilde e_R}\simeq
-3.2+\frac{12.4}{N_\Phi},\quad \delta_{\tilde d_R}-\delta_{\tilde l_L}
\simeq -2.5 + \frac{10.4}{N_\Phi},
\eea
for $M_\Phi={\cal O}(10)\,{\rm TeV}$.
In Fig$.$ \ref{DAM-F}, we consider more general situation with arbitrary
value of $M_\Phi$, and depict
$\delta_{\tilde{q}_L}-\delta_{\tilde{e}_R}$ and
$\delta_{\tilde{d}_R}-\delta_{\tilde{l}_L}$ for $2\leq N_\Phi\leq 5$.

In fact, some models of deflected anomaly mediation are severely constrained
by the condition to avoid tachyonic slepton, which typically requires a
large value of $N_\Phi$.
An example would be the model (\ref{effective1}) without the moduli $T_I$,
which gives $\beta=(n-1)/2$ and $M_\Phi\sim M_{Pl}(m_{3/2}/M_{Pl})^{1/(n-2)}$.
For the case of $n=4$, we need $N_\Phi\geq 10$ to avoid tachyonic slepton.
On the other hand, the corresponding $\delta_i$ are given by
\bea
\label{DAM-2}
&& \delta_{\tilde q_L}\simeq-0.74+\frac{5.6}{N_\Phi}, \quad
\delta_{\tilde{u}_R}\simeq -0.59+\frac{4.6}{N_\Phi},  \quad
\delta_{\tilde e_R} \simeq -0.10+\frac{0.6}{N_\Phi},
\nonumber \\
&& \delta_{\tilde d_R}\simeq -0.56+\frac{4.4}{N_\Phi},\quad
\delta_{\tilde l_L} \simeq -0.22 + \frac{1.44}{N_\Phi}.
\eea
For $N_\Phi=10$, which is the minimal value avoiding tachyonic slepton,
$\delta_i$ are all small, and then the model is difficult to be distinguished
from the mirage mediation with $\alpha=3/2$ and
$\tilde{m}_5^2=\tilde{m}_{10}^2=0$.
Similar situation occurs for the case that $X$ is stabilized by the radiative
correction to its K\"ahler potential, {\it e.g.} with $N_\Phi=10$,
$M_\Phi\sim 10^{12}\,{\rm GeV}$ and $\alpha=1$.

\section{Phenomenology of some examples}

In the previous section, we have examined generic feature of mass spectrum
of deflected mirage mediation, assuming the $SU(5)$ unification of matter
multiplets.
The effective supergravity models only containing the mass scales
$F^C/C\approx m_{3/2}$ and $M_{Pl/GUT}$, e.g. those of (\ref{effective1}) and
(\ref{GM-mass}), predict the relation (\ref{nd-scale}) with $\alpha\approx 1$.
In such models, the low energy mass spectrum of mirage mediation
($\alpha\approx 1$) is robust against the gauge threshold corrections.
On the other hand, once we arrange the special form of K\"ahler and super
potentials as in the model of (\ref{10TeV-DAM}), or introduce
(explicitly or dynamically) a new mass scale other than $M_{Pl/GUT}$ as in the
model of (\ref{ADS-potential}), the mass spectrum can dramatically change as
expected from Fig$.$ \ref{DMM-P}, although a realization of such models is
rather obscure in the string framework.
In the following, we discuss two phenomenological applications of deflected
mirage mediation representing these two cases.

\subsection{Accidental little SUSY hierarchy}

One of the virtues of the MSSM is the radiative electroweak symmetry breaking
\cite{Inoue:1982pi}.
The Higgs mass parameter, $m_{H_u}^2$ is automatically driven to negative due
to the renormalization group running by the top Yukawa coupling, even though
it is given a positive value at some high energy scale, $\Lambda$.
This radiative correction is controlled by the average stop mass,
$m_{\tilde{t}}^2$,
\bea
\delta m_{H_u}^2 \sim -\frac{3}{4\pi^2} y_t^2 m_{\tilde{t}}^2
\ln \left(\frac{\Lambda}{m_{\tilde{t}}} \right),
\eea
therefore, barring fine-tuning of the initial condition, we anticipate
$|m^2_{H_u}|\sim m^2_{\tilde{t}}$ for $\Lambda$ hierarchically larger than
$m_{\tilde{t}}$.
On the other hand, the lightest Higgs boson mass in the MSSM is approximated by
\bea
m_{h^0}^2 &\approx& M_Z^2 \cos 2\beta + \frac{3 y_t^2}{4 \pi^2} m_t^2
\ln\left( \frac{m_{\tilde{t}}^2}{m_t^2} \right),
\eea
where $\tan\beta=\langle H_u\rangle/\langle H_d\rangle$.
(Note that in the previous section $\beta$ has been used to parameterize the
anomaly to gauge mediation ratio.)
In order to fulfill the lower bound of the SM Higgs boson mass obtained at LEPII,
$m_{h_0} > 114\,{\rm GeV}$, we need the stop mass as heavy as
$m_{\tilde{t}} \gtrsim 600\,{\rm GeV}$.
Thus the Higgs mass parameter is generally expected to be
$|m_{H_u}| \gtrsim 600\,{\rm GeV}$.
While one of the conditions of the electroweak symmetry breaking tells us
\bea
\frac{M_Z^2}{2} &\approx& -m^2_{H_u} - |\mu|^2,
\eea
for $\tan\beta$ not too close to $1$. Here $\mu$ is the higgsino mass parameter
which does not break SUSY.
(Note that in the previous section $\mu$ has been used to parameterize the
renormalization point of running soft parameters.)
This means that we are forced to fine-tune the parameters, $m^2_{H_u}$ and
$|\mu|^2$ at less than $1\,$\% level to obtain the observed size of $M_Z$,
despite these two parameters are expected to be not correlated.
This not fatal but uncomfortable fine-tuning in the MSSM arising from the
hierarchy between the electroweak scale and the SUSY breaking mass scale is
called `the little SUSY hierarchy problem' \cite{Barbieri:1987fn}.

\begin{figure}[t]
\begin{center}
\begin{tabular}{c c}
\includegraphics[width=0.45\textwidth, clip]{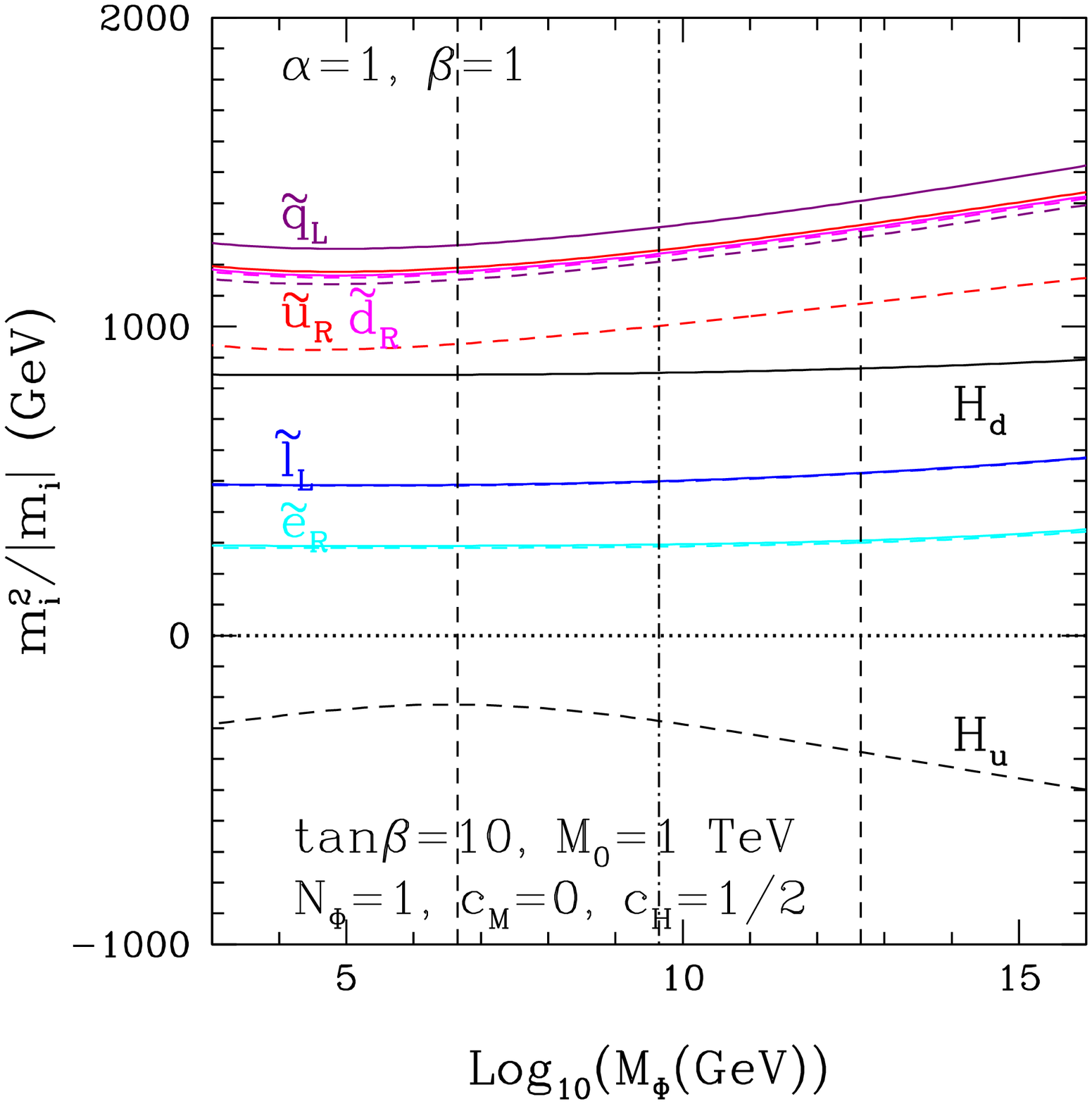}
&
\includegraphics[width=0.45\textwidth, clip]{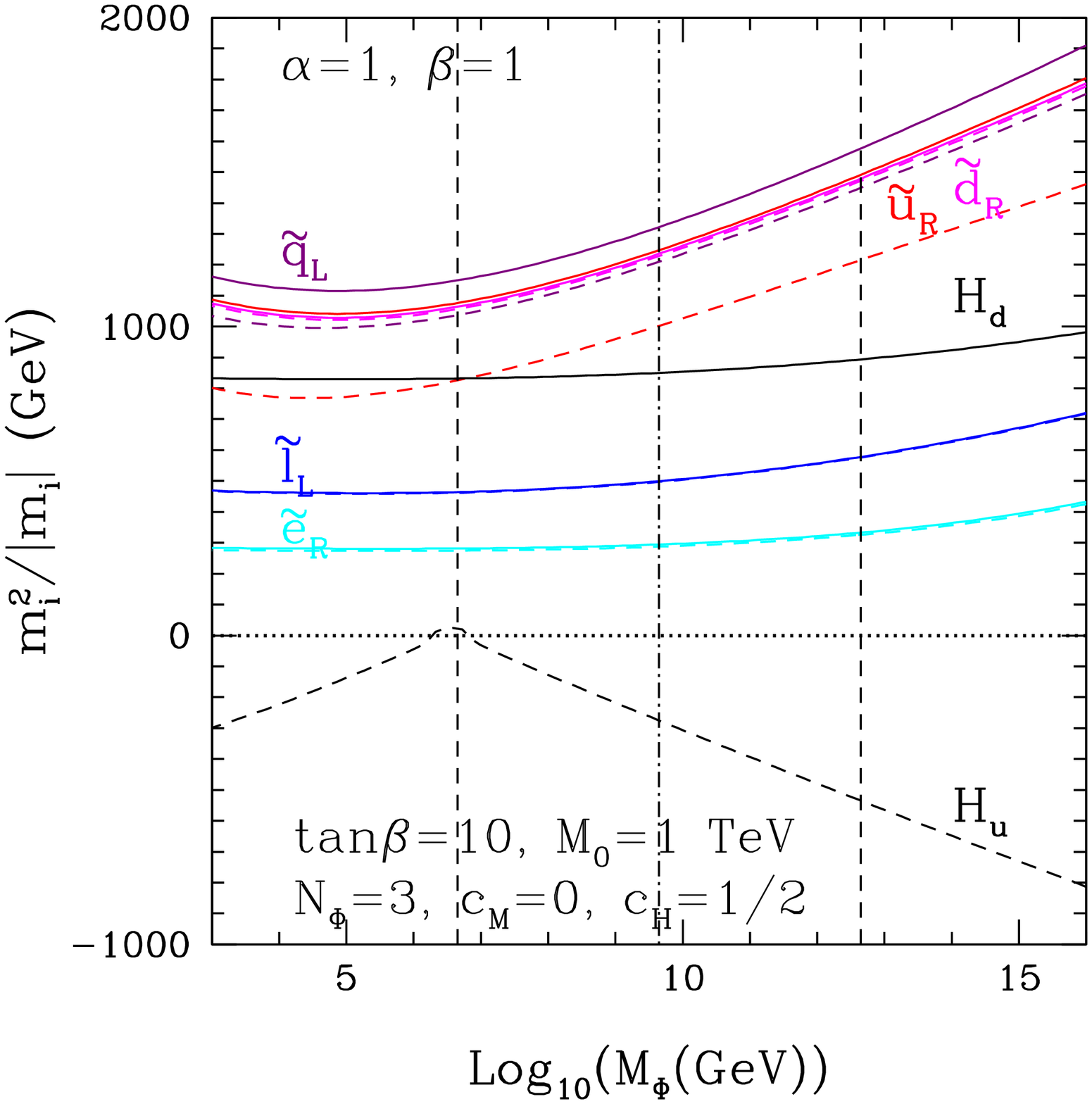}
\end{tabular}
\caption{Accidental little SUSY hierarchy in deflected mirage mediation
($\alpha=1$) for the case that $X$ is stabilized by the radiative effects in
K\"ahler potential.
The left panel shows the case for $N_\Phi=1$ while the right panel for
$N_\Phi=3$.
In all of them, the modular weights are chosen as
$c_M \equiv \tilde m^2_{\tilde{q}_L,\tilde{u}_R,
\tilde{d}_R,\tilde{l}_L,\tilde{e}_R}/M^2_0=0$
and $c_H\equiv \tilde m^2_{H_u, H_d}/M^2_0=1/2$.
Other SUSY parameters are set to $\tan\beta=10$ and $M_0=1\,{\rm TeV}$.
The vertical dashed lines indicate the predicted range of $M_\Phi$.
The vertical dot-dashed line represents the gauge threshold scale leading to
$R=1$.
}
\label{fig:little-hierarchy1}
\end{center}
\end{figure}

\begin{figure}[t]
\begin{center}
\begin{tabular}{c c}
\includegraphics[width=0.45\textwidth, clip]{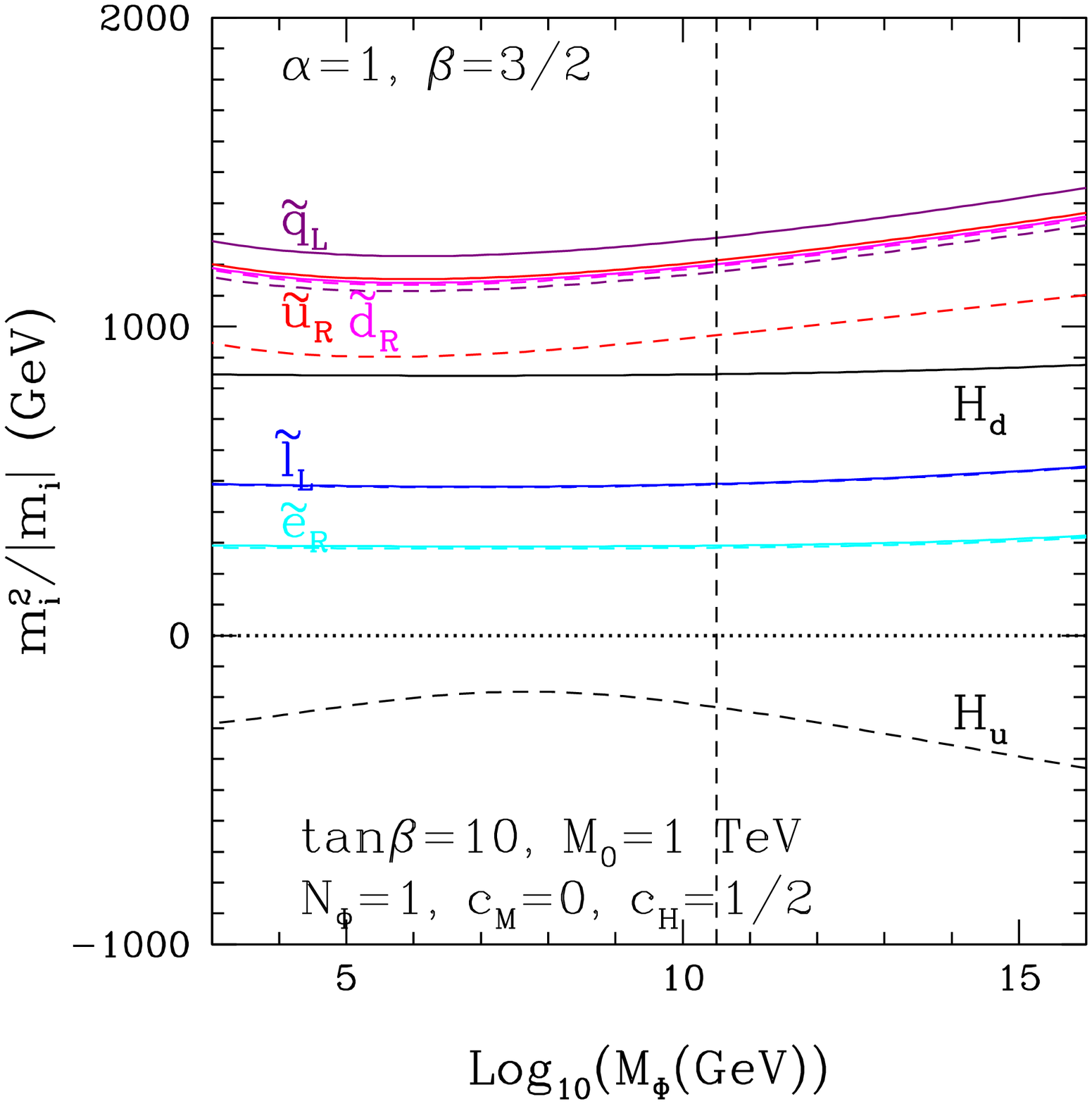}
&
\includegraphics[width=0.45\textwidth, clip]{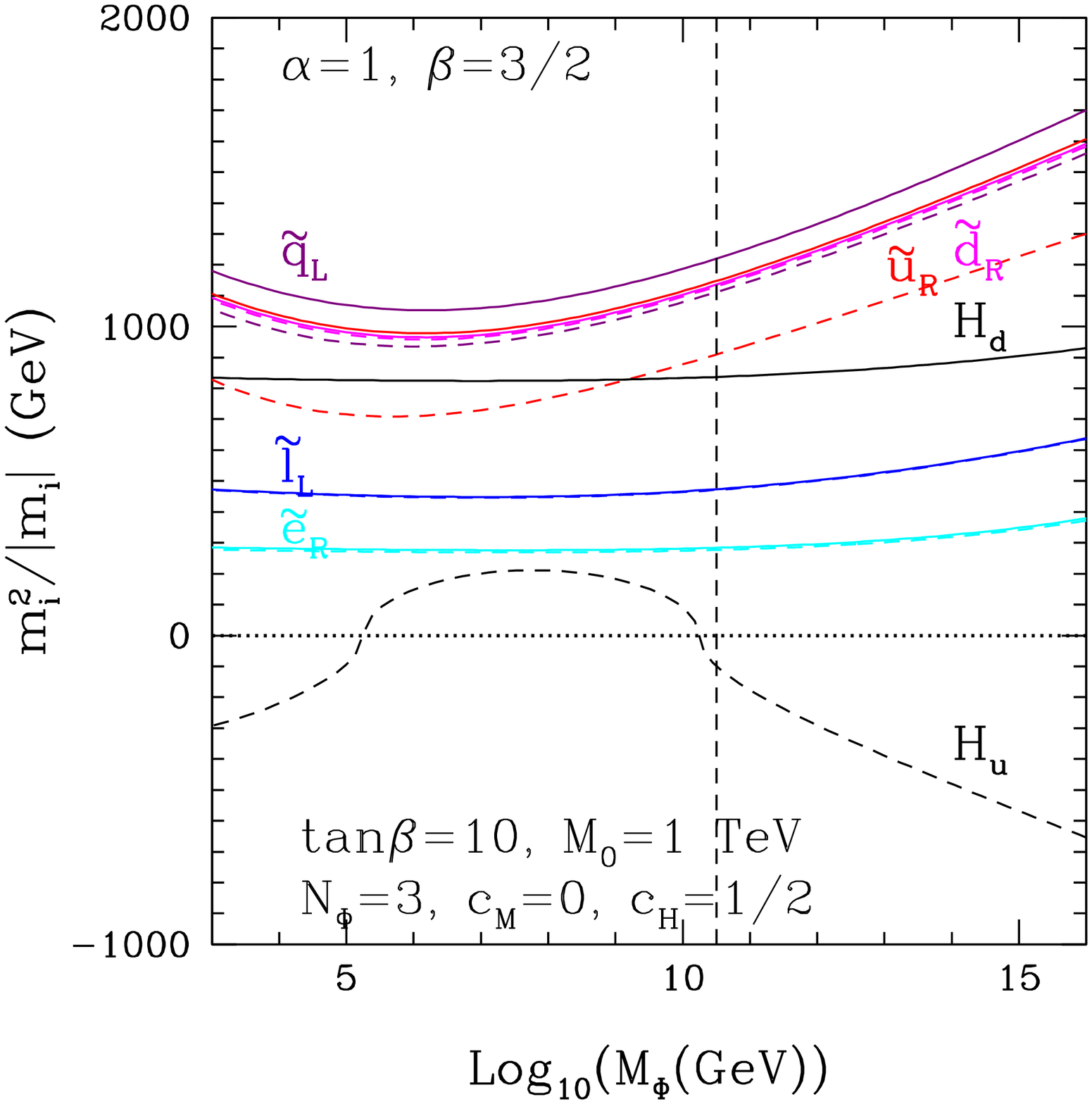}
\end{tabular}
\caption{Accidental little SUSY hierarchy in deflected mirage mediation
($\alpha=1$) for the case with $X$ stabilized by the higher dimensional operator
in superpotential.
SUSY parameters and modular weights are same as in Fig$.$
\ref{fig:little-hierarchy1}.
The vertical dashed line indicates the predicted value of $M_\Phi$.
}
\label{fig:little-hierarchy2}
\end{center}
\end{figure}

One obvious solution is having $m^2_{H_u}\sim M_Z^2 \ll m^2_{\tilde{t}}$ by
accident due to a choice of the boundary condition at $\Lambda$.
However, in mirage mediation, it is not apparent whether one can achieve such
a pattern or not, because the choice of the modular weights is discrete.
The moderate modification of the spectrum by the deflection may help to obtain
the desired mass pattern\footnote{For a different approach to this problem in
mirage mediation, see \cite{Choi:2005hd}.
It has also been argued that a negative stop mass-square at high
renormalization point can reduce the fine-tuning \cite{Dermisek:2006ey}.}.
In Fig$.$ \ref{fig:little-hierarchy1} and \ref{fig:little-hierarchy2}, we show
such an accidental little SUSY hierarchy achieved by the deflected mirage
mediation at $\alpha=1$, where we chose
$c_{\tilde{q}_L,\tilde{u}_R,\tilde{d}_R,\tilde{l}_L,\tilde{e}_R}
\equiv \tilde m^2_{\tilde{q}_L,\tilde{u}_R,\tilde{d}_R,\tilde{l}_L,
\tilde{e}_R}/M^2_0=0$,
$c_{H_u,H_d}\equiv \tilde m_{H_u,H_d}^2/M^2_0=1/2$ and $M_0=1\,{\rm TeV}$.
In Fig$.$ \ref{fig:little-hierarchy1}, we present the case that $X$ is stabilized
by the radiative effects in K\"ahler potential, while in Fig$.$
\ref{fig:little-hierarchy2}, the case for stabilization by the higher dimensional
operator in superpotential.
The left panels adopt $N_\Phi=1$ and the right ones $N_\Phi=3$.
In all of them, the dashed curves denote the 3rd generations.
In Fig$.$ \ref{fig:little-hierarchy1}, the dot-dashed line indicates
the gauge threshold scale at which the cancellation between gauge and moduli
mediations leads to $R=1$.
For $R=1$, neglecting the effects of Yukawa couplings, the sfermion masses
simply reduce to the values in pure mirage mediation.
The deviation associated with Yukawa couplings is also vanishing for $R=1$ if
the mirage condition (\ref{mirage_condition}) is satisfied for the
moduli-mediated soft terms.

For $N_\Phi=1$, the effect of deflection is limited.
However, for the case that $X$ is stabilized by the radiative effects, we can
obtain an improved hierarchy for $N_\Phi=3$ and $M_\Phi \sim 10^{6-7}\,{\rm GeV}$,
which is within the plausible range of this stabilization mechanism indicated
by the vertical dashed lines \cite{Nakamura:2008ey}.
In this case we need to break the PQ symmetry slightly to make
the axion heavy so that it will not be produced in astrophysical processes
and evade the bound, $10^{9}\,{\rm GeV}\,\lesssim M_\Phi$ coming from the burst
duration of supernova SN1987A and the cooling of globular-cluster stars and
white dwarf \cite{Amsler:2008zz}.
In Fig$.$ \ref{fig:little-hierarchy2}, which is for the case with $X$ stabilized by
the non-renormalizable superpotential term, the vertical dashed line shows
the gauge threshold scale predicted for $n=4$ with $x=1$ in
(\ref{higher-dim-operator}).
Again, we have an improved hierarchy for $N_\Phi=3$.
In both cases, we need a mechanism to generate $\mu$ and $B\mu$ terms of
appropriate size, which does not disturb the mass spectrum.
For instance, in the first case we can employ the mechanism described in
\cite{Pomarol:1999ie, Nakamura:2008ey} in weak coupling limit, which employs a
term $(X^\dag/X)\, H_d H_u +{\rm h.c.}$ in the K\"ahler potential to generate
the desired values of $\mu$ and $B\mu$.
In the second case, we can use the same mechanism by introducing another
singlet $X^\prime$ stabilized at $M_{\rm mir}$ by the K\"ahler potential, which
minimizes the deflection (see (\ref{nd-scale})).

\subsection{Gluino lightest supersymmetric particle in deflected mirage mediation}

\begin{figure}[t]
\begin{center}
\begin{tabular}{c c}
\includegraphics[width=0.45\textwidth, clip]{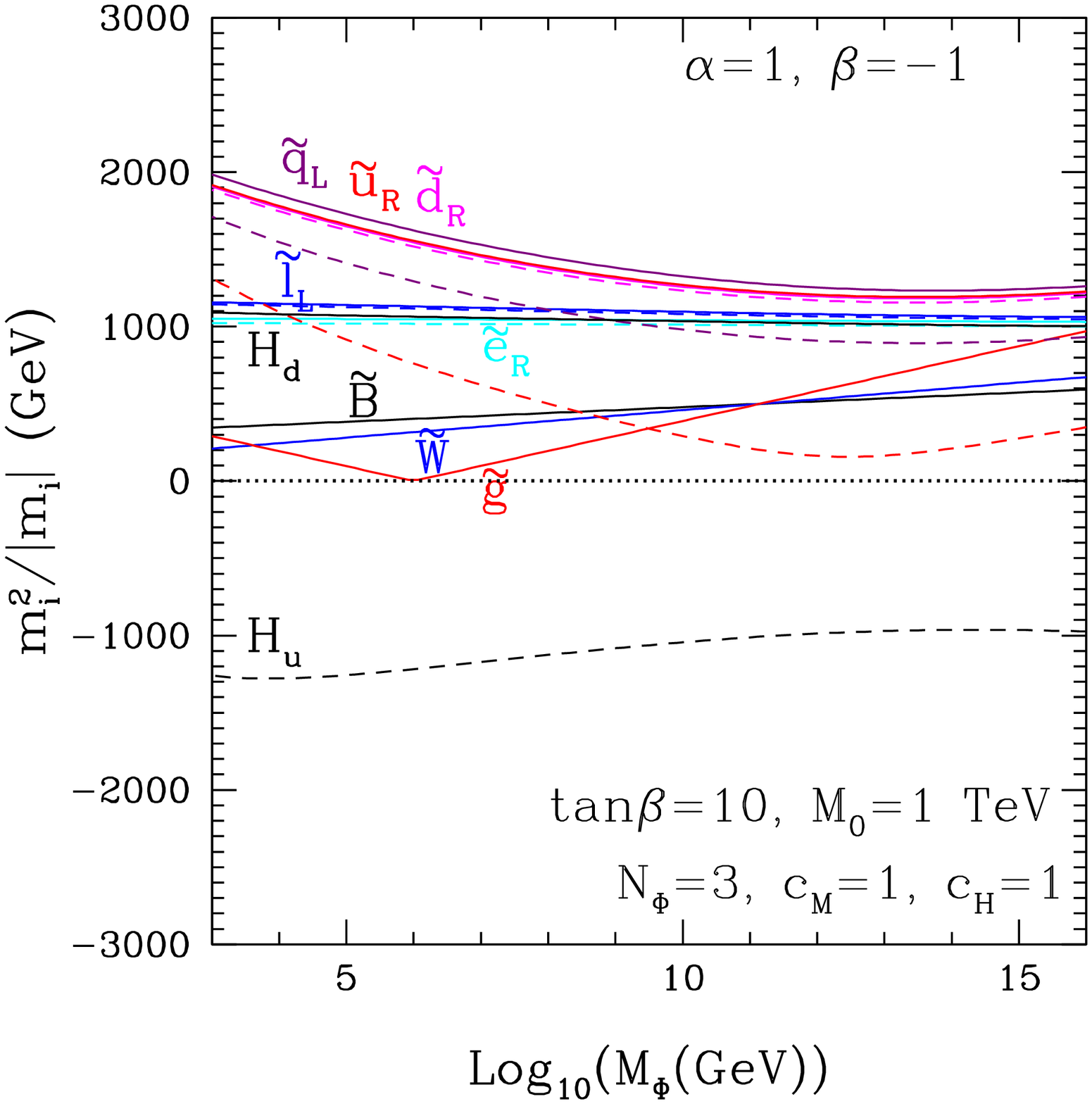}
&
\includegraphics[width=0.45\textwidth, clip]{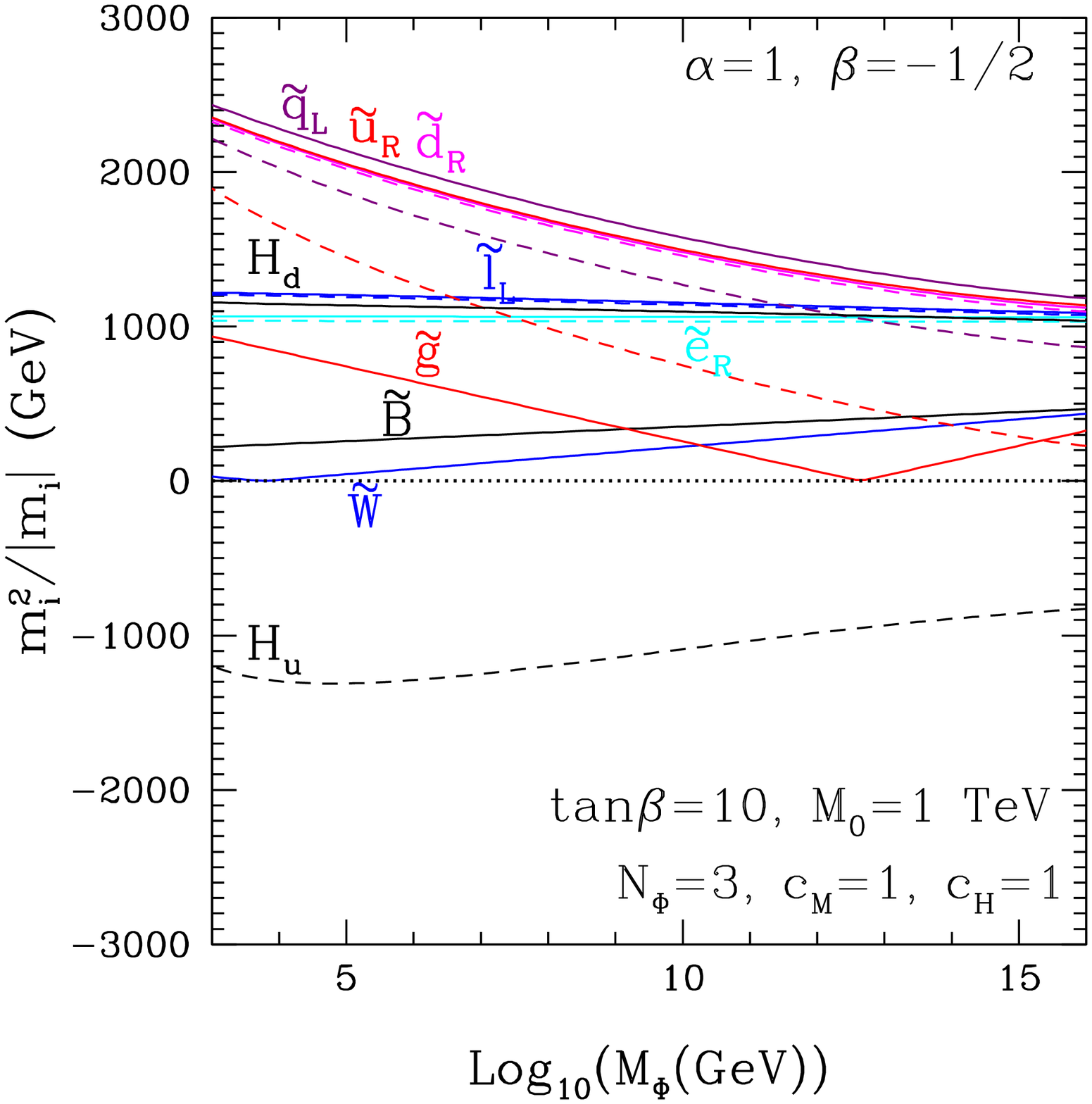}
\end{tabular}
\caption{Gluino and wino LSP in defected mirage mediation ($\alpha=1$).
The dashed curves indicate the 3rd generation.
The left panel shows the case for $\beta=-1$ and the right for $\beta=-1/2$.
All the modular weights are chosen so that $c_i\equiv \tilde m_i^2/M^2_0=1$.
Other parameters are chosen as $\tan\beta=10$, $M_0=1\,{\rm TeV}$ and $N_\Phi=3$.
}
\label{fig:gluinoLSP}
\end{center}
\end{figure}

As we have seen in the previous section, we can considerably reduce $R$
(or increase $\alpha_{\rm eff}$) from $1$ if we chose the special form of K\"ahler and
superpotential (\ref{10TeV-DAM}) \cite{Hsieh:2006ig} or use the negative power
superpotential for $X$ (\ref{ADS-potential}) \cite{Okada:2002mv, Everett:2008qy}.
In such a case, we have possibilities that the lightest supersymmetric particle
(LSP) becomes gluino ($\alpha_{\rm eff} \approx 3$) or wino
($\alpha_{\rm eff} \gg 1$) as shown in Fig$.$ 1 of \cite{Choi:2005uz}.
On the other hand, in contrast to mirage mediation, $(\tilde{m}^{\rm eff}_i)^2$
does not vanish in the limit $M^{\rm eff}_0\, (R) \to 0$ as seen in
(\ref{sfermion_def2}).
This considerably dilutes the renormalization scale dependent part in
(\ref{sfermion_def}) relative to the constant term, which leads to the
`quasi-infrared-fixed-point' behavior of the scalar mass as first observed in
\cite{Everett:2008qy}.
This effect makes squarks and sleptons somewhat heavier than the gauginos for
small $R$ ($M_\Phi$) as shown in Fig$.$ \ref{fig:gluinoLSP}.
Therefore a direct production of the squarks or sleptons at hadron collider is
suppressed.
Their on-shell states also can not appear in the cascade decay of gluino
in the wino LSP case.
As the physics of the wino LSP has been extensively examined in association with
the anomaly mediation \cite{winolsp1, winolsp2}, here we will focus on the more
exotic case: the gluino LSP.

Phenomenology of the gluino LSP or meta-stable gluino has been investigated from
various motivations
\cite{Baer:1998pg, Raby:1998xr, Kraan:2004tz, Bussey:2006vx, Fairbairn:2006gg},
particularly in the context of split supersymmetry recently
\cite{ArkaniHamed:2004fb, Kilian:2004uj, Hewett:2004nw, Cheung:2004ad,
Toharia:2005gm, Arvanitaki:2005nq}.
Most of the analyses can be directly applied to our case.
We summarize some of the results below.

Since squarks are somewhat heavier than gluino in the present scenario, gluino is
mainly produced via the processes,
$q \bar{q} \to g^\ast \to \tilde{g} \tilde{g}$ or $g g \to g^\ast \to \tilde{g}
\tilde{g}$ at the hadron collider such as Tevatron and LHC.
While at the lepton collider it is always produced in association with quarks,
$e \bar{e} \to q \bar{q} \to q \bar{q} \tilde{g} \tilde{g}$ since the leptons
are color singlets.
The produced gluino hadronizes into a color-singlet composite state called
R-hadron \cite{Farrar:1978xj}.
The bound states of gluino and color-octet hadron,
$\tilde{g}g$, $\tilde{g}q\bar{q}$ and $\tilde{g}qqq$ are known as R-gluon,
R-meson and R-baryon, respectively.
Phenomenologically, the most relevant question is the stability of the charged
particles which leave tracks in the detector.
It depends on the identity of the lightest R-hadron and their mass differences.
The mass spectrum of R-hadron is estimated by the MIT bag model
\cite{Chanowitz:1983ci} and the quenched lattice simulation \cite{Foster:1998wu}
which predict iso-triplet vector R-meson and R-gluon ($J^{PC}=1^{+-}$) as the
lightest R-hadron, respectively.
In both cases their mass difference is smaller than the pion mass.
Thus the vector R-meson is stable against hadronic decay.
The mass difference of the lightest R-baryon and the ground state of R-gluon or
R-meson is also estimated to be smaller than the nucleon mass and the R-baryon
is stable \cite{Chanowitz:1983ci, Foster:1998wu, Kraan:2004tz}.
Therefore the probability for gluino to hadronize into the charged states,
$P$ is expected to be non-negligible, although the reliability of such a
conclusion is limited by those of the calculation method and adopted
assumptions.

Inside the detector the R-hadron deposits energy via hadronic interaction
with nucleus and, if it is charged, ionization of the detector material.
The gluino is typically produced with momentum similar order of its mass
($m_R \sim 100\,{\rm GeV}$) and thus the R-hadron is relativistic but slow,
$\beta < 1$.
The energy involved in the hadronic interaction is approximated by
$Q=\sqrt{s}-m_R -m_N\approx (\gamma-1) m_N$ with $\gamma = 1/\sqrt{1-\beta^2}$
\cite{Kraan:2004tz}.
Therefore the interaction is soft nevertheless the energy carried by the
R-hadron is huge ($\sim 100\,{\rm GeV}$).
Then the neutral R-hadron traverses the detector repeatedly kicking off the
nucleon inside the nucleus and the soft secondary particles dissipating a small
fraction of its kinetic energy at each collision.
Eventually it penetrates the detector carrying away significant amount of
missing energy.
This behavior is in contrast to the ordinary hadron which develops shower and
exponentially dumps its energy in the calorimeter.
The ionization energy loss of the charged R-hadron is calculated by the standard
Bethe-Bloch formula \cite{Amsler:2008zz}.
Since the energy loss, $-d E/dx$ is proportional to the inverse of $\beta^2$
while the average hadronic energy loss per collision behaves like
$\langle \Delta E \rangle \propto \gamma$, the ionization plays a minor role for
large $\beta$ ($\gtrsim 0.9$), however, it quickly dominates over the hadronic
interaction as the R-hadron is slowed down \cite{Hewett:2004nw}.
The heavily ionizing track provides a characteristic signal of the slowly moving
charged R-hadron.
It is noted that even if the R-hadron is neutral it can change its identity
in the hadronic interaction.
If the neutral and charged states are both stable, it transforms from neutral
to charged and vice versa along with its path (flipper scenario).

Based on the careful inference about the nature of R-hadron as explained above,
the pioneering work by \cite{Baer:1998pg} excluded the gluino mass range,
$3 \, {\rm GeV}\, \leq m_{\tilde{g}}\leq 130\,{\rm GeV}$ at $95\,$\% CL almost
independent of $P$ using jet+$\cancel{E}$ channel in LEP and Tevatron CDF RUN I,
while $50\,{\rm GeV}\, \leq m_{\tilde{g}}\leq 200\,{\rm GeV}$ for $P\geq 1/2$
by a heavily-ionizing track search in Tevatron CDF.
On the other hand, the analysis in \cite{Hewett:2004nw} emphasized a model
independent role of the high $p_T$ monojet which is produced in association
with the gluino pair.
They obtained a conservative bound,
$m_{\tilde{g}}\geq 170\,{\rm GeV}$, independent of $P$ using the Tevatron Run I
data and projected it to $m_{\tilde{g}}\geq 210\,{\rm GeV}$ for Run II and
$m_{\tilde{g}}\geq 1.1\,{\rm TeV}$ for LHC.
They also estimated a reach of the charged track search ($P=1$) as
$m_{\tilde{g}} = 270\,{\rm GeV}$ for CDF Run I with integrated luminosity
$100\, {\rm  ps}^{-1}$, $m_{\tilde{g}}= 430\,{\rm GeV}$ for Run II with
$2\,{\rm fb}^{-1}$ and $m_{\tilde{g}}= 2.4\,{\rm TeV}$ for LHC with
$100\,{\rm fb}^{-1}$.
More serious study on the R-hadron discovery potential in LHC ATLAS detector
has been performed using the ATLAS fast simulation framework \cite{Kraan:2004tz}.
They have taken into account the flipper behavior of the R-hadron and perform Monte
Carlo simulation based on GEANT3.
Using event selection with global event variables such as the missing
transverse energy $\cancel{E}_T$, the total sum of the transverse energy
$E_T^{\rm tot}$ and the transverse momentum ($p_T$) measured in the muon
chamber, they concluded that the R-hadron is discovered for the mass up to
$1.4\,{\rm TeV}$ with the integrated luminosity $30\,{\rm fm}^{-1}$.
While the reach extends to $1.7\,{\rm TeV}$ if the time-of-flight information
for slow-moving R-hadron between the muon chambers is used in the event
selection in stead of $\cancel{E}_T$ and $E_T^{\rm tot}$.
From these studies we can conclude that still the gluino LSP scenario of our
interest has a considerable portion of parameter space, however, it will be
fully examined in the relatively early stage of LHC.

The stable R-hadron in cosmological time scale having electric or hadronic
interaction will conflict with various phenomenological constraints such as
heavy isotope searches \cite{Smith:1982qu, Hemmick:1989ns, Javorsek:2001yv}
unless the relic abundance is sufficiently small
\cite{Mohapatra:1998nd, JavorsekII, Baer:1998pg}.
An alternative interesting possibility is that the gluino is unstable but
decays outside of the detector as in the split supersymmetry, which circumvents
the cosmological difficulty but will not change the collider signature we have
discussed above.
A concrete example is given by the axionic extension of deflected mirage
mediation as discussed in \cite{Nakamura:2008ey}.
Because of the recursive nature of the mass spectrum of deflected mirage
mediation with multi-thresholds, we can introduce the axion superfield
$S$ stabilized by the K\"ahler potential in addition to $X$ without disturbing
the low energy mass spectrum as long as $\langle X \rangle <\langle S \rangle
\approx M_{\rm mir}$.
The LSP is now axino, $\tilde{a}$ whose mass is one-loop suppressed relative
to the gauginos since there is no tree-level contribution from the
superpotential.
The next lightest supersymmetric particle (NLSP) is gluino, which decays to
the axino via $\langle S \rangle$-dependence in the gauge coupling constant
introduced by the threshold correction.
The corresponding interaction Lagrangian is given by \cite{Covi:1999ty}
\bea
{\cal L}_{\tilde{g}\tilde{a} g} &=& \frac{\alpha_s N_\Phi}{16\sqrt{2}\pi}
\frac{1}{\langle S \rangle} \bar{\tilde{a}}\gamma_5 \sigma^{\mu \nu}
\tilde{g} g_{\mu \nu} + {\rm h.c.},
\eea
which yields the decay width
\bea
\Gamma(\tilde{g} \to \tilde{a} g)
&\simeq& \frac{\alpha_s^2 N_\Phi^2}{32\pi^3}
\frac{m_{\tilde{g}}^3}{\langle S \rangle^2},
\eea
and the life time
\bea
\label{eq:gluino-decay}
\tau_{\tilde{g}} &=& \Gamma(\tilde{g} \to \tilde{a} g)^{-1}
\simeq 5.7\times 10^{-7}\,{\rm sec}\,
\left(\frac{m_{\tilde{g}}}{200\,{\rm GeV}} \right)^{-3}
\left(\frac{\langle S \rangle}{10^{10} \,{\rm GeV}}
\right)^{2} N_\Phi^{-2},
\eea
where $N_\Phi$ is the number of messengers coupling with the axion.
For $m_{\tilde{g}}\lesssim 1\,{\rm TeV}$ and $N_\Phi \simeq 1$, most of
the decay occurs outside of the detector and the discovery prospects
discussed above is applicable.
On the other hand, for $m_{\tilde{g}}\gtrsim \,{\rm TeV}$ or $N_\Phi \gg 1$,
gluinos decay inside the detector, which results in a displaced jet vertex
with missing energy.
It would provide a clear signature and a similar experimental reach as in the
case of the heavily ionizing track, although a detailed study in a realistic
circumstance is mandatory for deriving any definitive conclusion.

The relic axino abundance in deflected mirage mediation with the gluino NLSP
can be estimated by following the discussion in \cite{Nakamura:2008ey}.
As is well known, if we have a light modulus $T$ as in (deflected) mirage
mediation, its coherent oscillation dominates the energy density of the universe
after the inflation \cite{de Carlos:1993jw}.
Eventually it decays and  produced entropy dilutes everything existed before.
The saxion also oscillates, but it is harmless as long as
$\langle S \rangle \lesssim 10^{11}\,{\rm GeV}$ since it decays faster than the
modulus\footnote{If the saxion oscillation dominates the universe, axions
produced from its decay upset the successful prediction of Big Bang
nucleosynthesis.
Therefore axion can not be a dominant component of the dark matter in this
scenario \cite{Nakamura:2008ey}.}.
The reheating temperature of the modulus is estimated as \cite{Nakamura:2008ey}
\bea
T_R &=& \left(\frac{90}{\pi^2 g_\ast(T_R)}\right)^{1/4}\sqrt{M_{Pl} \Gamma_T}
\simeq 0.15\, {\rm GeV} \left(\frac{g_\ast(T_R)}{10}\right)^{-1/4}
\left( \frac{m_T}{10^6\, {\rm GeV}} \right)^{3/2} d_g,
\eea
where $g_\ast(T_R)$ denotes the effective bosonic degrees of freedom at $T_R$
for the energy density and $d_g$ is a model dependent parameter of order unity
defined as $d_g \equiv 2 (-3{\partial_T\partial_{T^*}\ln\Omega_{\rm mod}})^{-1/2}
\partial_T \ln({\rm Re}(\tilde{f}_a))$.
It is marginally smaller than the thermal decoupling temperature of gluino
\cite{Baer:1998pg}
\bea
T_F &=& m_{\tilde{g}}/x_F \approx 6 \,{\rm GeV}
\left( \frac{m_{\tilde{g}}}{200\,{\rm GeV}}\right).
\eea
Therefore the thermal relic abundance of gluino is not suitable to calculate
the axino abundance.
Actually, the dominant contribution to the axino relic abundance comes from
the gravitino which is produced by the decay of the modulus
\cite{Nakamura:2006uc}.
The gravitino decay eventually produces at least one gluino (or axino) and
hence one axino.
The gravitino yield which is conserved in the adiabatic expansion of the
universe is given by \cite{Nakamura:2006uc}
\bea
\label{eq:gravitino-yield}
Y_{3/2} &\equiv& \frac{n_{3/2}}{s} \,\simeq\,
\frac{3}{2} \frac{T_R}{m_T} B_{3/2}^T \sim 4.3 \times 10^{-9}
\left(\frac{g_\ast(T_R)}{10}\right)^{-1/4}
\left( \frac{m_T}{10^6\, {\rm GeV}} \right)^{1/2} d_g.
\eea
The gluino decay to axino potentially competes with the gluino annihilation.
The perturbative annihilation cross section in the zero relative velocity
limit ($\beta=0$) is given by
$\sigma_{\rm ann}\beta \simeq (171\pi\alpha_s^2/64 m_{\tilde{g}}^2)$
\cite{Baer:1998pg}.
Then the inverse of annihilation rate is estimated as
\bea
\Gamma_{\rm ann}^{-1} &\simeq&
\left(n_{3/2}\, \sigma_{\rm ann}\, \beta \right)^{-1}
\simeq
\left( Y_{3/2}\, s(T_{3/2}) \sigma_{\rm ann}\beta\right)^{-1}
\nonumber \\
&\simeq& 1.6 \times 10^{-5} \,{\rm sec}
\left( \frac{m_T}{10^6\, {\rm GeV}} \right)^{-1/2}
\left( \frac{m_{3/2}}{10^5\, {\rm GeV}} \right)^{-9/2}
\left( \frac{m_{\tilde{g}}}{200\, {\rm GeV}} \right)^{2} d_g^{-1},
\eea
which is considerably larger than (\ref{eq:gluino-decay}).
Thus the annihilation is negligible and (\ref{eq:gravitino-yield}) gives
a good estimation of the axino yield. Then the current axino density is given
by
\bea
\Omega_{\tilde{a}}h^2 &=& \frac{Y_{3/2} s_0 h^2 }{3M_P^2 H_0^2} \simeq 1.2
\left(\frac{m_{\tilde{a}}}{1\,{\rm GeV}}\right)
\left(\frac{g_\ast(T_R)}{10} \right)^{-1/4}
\left(\frac{m_T}{10^6\,{\rm GeV}}\right)^{1/2} d_g,
\eea
where $s_0$ and $H_0$ are the entropy density and Hubble constant of the
current universe, respectively.
Thus the axino, which has one-loop suppressed mass relative to the gaugino
($\sim 100\,{\rm MeV}$), accompanied with the gluino NLSP can naturally saturate
the observed dark matter density as discussed in \cite{Nakamura:2008ey} for
other NLSPs.

\section{Conclusion}

In this paper, we have discussed the sparticle mass pattern in deflected mirage
mediation scenario of supersymmetry breaking, in which all of the three known
flavor conserving mediations, i.e. dilaton/moduli, anomaly and gauge mediations,
contribute to the MSSM soft parameters.
Starting with a class of string-motivated effective supergravity models that
realize deflected mirage mediation, we analyzed the renormalization group running
of soft parameters to derive the (approximate) analytic expression of low energy
sparticle masses at the TeV scale.
We also discussed more detailed phenomenology of two specific examples, one with
an accidental little hierarchy between $m^2_{H_u}$ and other soft mass-squares
and another with gluino NLSP, that can be obtained  within deflected mirage
mediation scenario.

If some sparticles masses are in the sub-TeV range, the corresponding sparticles
can be copiously produced at the CERN LHC, and then one may be able to measure
their masses with the methods proposed in \cite{massmeasurement}.
Our results then can be used to interpret the experimentally measured sparticle
masses within the framework of the most general flavor and CP conserving mediation
scheme.

\vskip 1cm {\bf Acknowledgement}
\vskip 0.5cm
KC is supported by the KRF Grant funded by the Korean Government
(KRF-2005-210-C0006 and KRF-2007-341-C00010) and the BK21 program of Ministry
of Education.
MY and SN are partially supported by the grants-in-aid from the Ministry of
Education, Culture, Sports, Science and Technology (MEXT) of Japan (No.16081202 and
No.17340062) and the global COE program ``Weaving Science Web beyond
Particle-Matter Hierarchy''.
KO is supported by the Grant-in-aid for Scientific Research No.18071007 and
No.19740144 from MEXT of Japan.
KO also thanks Yukawa Institute in Kyoto University for the use of Altix3700 BX2.

\vskip 1cm

{\bf Appendix:}
\vskip 0.5cm

In this appendix, we  present the analytic expression of sfermion soft parameters
including the effects of the top quark Yukawa coupling $y_t$, when the
moduli-mediated soft parameters of $Q_i=H_u,q_3,u_3$ at $M_{GUT}$ do {\it not}
satisfy the mirage condition (\ref{mirage_condition}).
As the expression for generic deflected mirage mediation is too much involved, so
not useful, here we present only the result in the mirage mediation limit without
gauge mediation contribution.

In the low $\tan\beta$ regime of the MSSM, neglecting the Yukawa couplings other
than $y_t$, the anomalous dimension of $Q_i$ reads
\bea
\gamma_i(\mu) &=& 2\sum_a C^a_i(Q_i)g^2_a(\mu)+k_i y^2_t(\mu),
\eea
where $k_i$ is non-vanishing only for $Q_i=H_u, q_3, u_3$ having the top Yukawa
interaction
\bea
k_{H_u} = -3, \quad k_{q_3} = -1, \quad k_{u_3} = -2.
\eea
Above the gauge threshold scale $M_\Phi$, the running top Yukawa coupling is
given by
\bea
y^2_t(\mu) &=& \frac{y^2_t(M_{GUT})\dot G(\mu)}{1-\frac{3}{4\pi^2}
y^2_t(M_{GUT})G(\mu)}, \eea where \bea G(\mu) &=& \int^\mu_{M_{GUT}}
\frac{d\mu^\prime}{\mu^\prime}
\prod_a\left( 1 + \frac{b^H_a}{8\pi^2} g^2_{GUT}
\ln\left(\frac{M_{GUT}}{\mu^\prime}\right) \right)^{2C^a_t/b^H_a},
\eea
with $C^a_t=C^a_2(H_u)+C^a_2(q_3)+C^a_2(u_3)$.
Using this, we can find the analytic expressions for the sfermion soft
parameters even when the mirage condition (\ref{mirage_condition}) is not
satisfied:
\bea
\label{Yt-solution}
A_{ijk}(\mu) &=& \tilde A_{ijk}
- (k_i+k_j+k_k)(\tilde A_{H_uq_3u_3}-M_0)\rho(\mu)
\nonumber \\
&&
-\, \frac{M_0}{8\pi^2}\Big(\gamma_i(\mu)+\gamma_j(\mu)+\gamma_k(\mu)\Big)
\ln\left(\frac{\mu}{M_{\rm mir}}\right),
\nonumber \\
m^2_i(\mu) &=& \tilde m^2_i - k_i\left[ (\tilde A_{H_uq_3u_3}
-M_0)^2\Big(1+6\rho(\mu)\Big) + (\tilde m^2_{H_u} +\tilde
m^2_{\tilde{t}_L} + \tilde m^2_{\tilde{t}_R} -M^2_0)
\right]\rho(\mu)
\nonumber \\
&& -\, \frac{M_0}{4\pi^2}\left[ M_0\gamma_i(\mu) + k_i(\tilde
A_{H_uq_3u_3} -M_0)\Big(1+6\rho(\mu)\Big)y^2_t(\mu) \right]
\ln\left(\frac{\mu}{M_{\rm mir}}\right)
\nonumber \\
&&
-\, \frac{M^2_0}{8\pi^2}\dot\gamma_i(\mu)\left[
 \ln\left(\frac{\mu}{M_{\rm mir}}\right) \right]^2,
\eea where  \bea \rho(\mu) =
\frac{y^2_t(\mu)}{8\pi^2}\frac{G(\mu)}{\dot G(\mu)}.
\nonumber
\eea
It is obvious that the above solutions reduce to (\ref{eq:mirage1}) when the
mirage condition (\ref{mirage_condition}) is satisfied.

\end{document}